\documentclass[sigconf]{acmart}

\renewcommand\footnotetextcopyrightpermission[1]{}

\usepackage[english]{babel} 
\usepackage{tikz}
\usepackage{xurl}
\usepackage{amsmath}
\usepackage{bbm}
\usepackage{amssymb}
\usepackage{mathtools}
\usepackage{pifont}
\usepackage{algorithm}
\usepackage{algorithmicx}
\usepackage{algpseudocode}
\usepackage{amsmath}
\usepackage{amsthm}
\usepackage{microtype}
\usepackage{graphicx}
\usepackage{wasysym}
\usepackage{soul}

\usepackage{bbding}
\usepackage{subfigure}
\usepackage{booktabs} 
\usepackage{balance}
\usepackage{multirow}
\usepackage{multicol}
\usepackage{hyperref}
\usepackage[all]{nowidow}
\usepackage{color, colortbl} 

\def\denseitems{
  \itemsep1pt plus1pt minus1pt
  \parsep0pt plus0pt
  \parskip0pt\topsep0pt}

\newcommand*\circled[1]{\tikz[baseline=(char.base)]{
            \node[shape=circle,fill,inner sep=1pt] (char) {\textcolor{white}{#1}};}}

\newcommand{\hytt}[1]{\texttt{\hyphenchar\font=\defaulthyphenchar #1}}

\definecolor{Gray}{gray}{0.9}

\copyrightyear{2025}
\acmYear{2025}
\setcopyright{acmlicensed}\acmConference[CCS '25]{Proceedings of the 2025 ACM SIGSAC Conference on Computer and Communications Security}{October 13--17, 2025}{}
\acmBooktitle{Proceedings of the 2025 ACM SIGSAC Conference on Computer and Communications Security (CCS '25), October 13--17, 2025}
\acmDOI{10.1145/3719027.3744829}
\acmISBN{979-8-4007-1525-9/2025/10}

\begin{document}

\title{Analyzing PDFs like Binaries: Adversarially Robust PDF Malware Analysis via Intermediate Representation and Language Model}


\author{Side Liu}
\authornote{The Key Laboratory of Aerospace Information Security and Trusted Computing, Ministry of Education, School of Cyber Science and Engineering, Wuhan University.}
\affiliation{%
  \institution{Wuhan University}
  \city{Wuhan}
  \country{China}
}
\email{sidelau@whu.edu.cn}

\author{Jiang Ming}
\authornote{Department of Computer Science, School of Science and Engineering, Tulane University.}
\affiliation{%
  \institution{Tulane University}
  \city{New Orleans}
  \country{USA}
}
\email{jming@tulane.edu}

\author{Guodong Zhou}
\authornotemark[1]
\affiliation{%
  \institution{Wuhan University}
  \city{Wuhan}
  \country{China}
}
\email{zhouguodong@whu.edu.cn}

\author{Xinyi Liu}
\authornotemark[1]
\affiliation{%
  \institution{Wuhan University}
  \city{Wuhan}
  \country{China}
}
\email{xinyiliu@whu.edu.cn}

\author{Jianming Fu}
\authornotemark[1]
\affiliation{%
  \institution{Wuhan University}
  \city{Wuhan}
  \country{China}
}
\email{jmfu@whu.edu.cn}

\author{Guojun Peng}
\authornotemark[1]
\authornote{Corresponding author.}
\affiliation{%
  \institution{Wuhan University}
  \city{Wuhan}
  \country{China}
}
\email{guojpeng@whu.edu.cn}

\thanks{To appear in the Proceedings of The ACM Conference on Computer and Communications Security (CCS), 2025}




\renewcommand{\shortauthors}{Side Liu et al.}

\begin{abstract}

Malicious PDF files have emerged as a persistent threat and become a popular attack vector in web-based attacks. While machine learning-based PDF malware classifiers have shown promise, these classifiers are often susceptible to adversarial attacks, undermining their reliability.
To address this issue, recent studies have aimed to enhance the robustness of PDF classifiers. Despite these efforts, the feature engineering underlying these studies remains outdated. Consequently, even with the application of cutting-edge machine learning techniques, these approaches fail to fundamentally resolve the issue of feature instability.

To tackle this, we propose a novel approach for PDF feature extraction and PDF malware detection.
We introduce the PDFObj IR (PDF Object Intermediate Representation), an assembly-like language framework for PDF objects, from which we extract semantic features using a pretrained language model. Additionally, we construct an Object Reference Graph to capture structural features, drawing inspiration from program analysis. This dual approach enables us to analyze and detect PDF malware based on both semantic and structural features.
Experimental results demonstrate that our proposed classifier achieves strong adversarial robustness while maintaining an exceptionally low false positive rate of only 0.07\% on baseline dataset compared to state-of-the-art PDF malware classifiers.

\end{abstract} 

\maketitle

\section{Introduction}
\label{introduction}

The Portable Document Format (PDF) is among the most widely used file formats on the web~\cite{pdfa_popularity}, making it an attractive target for cybercriminals due to its ubiquity and versatility~\cite{2023Unit42Report,macfee_pdf,trustwave_pdf,PDFs_Why}. 
PDFs now dominate as the most commonly used malicious attachments in phishing campaigns, with nearly 70\% of these emails evading network-based defenses and 15\% bypassing endpoint security measures~\cite{Phishing2022cisa}.
Moreover, the increasing prevalence of cloud-based collaboration and remote work has led to the widespread integration of PDF readers within modern browsers, further heightening the threat posed by malicious PDFs. 
Attackers can exploit vulnerabilities in web applications, such as cross-site scripting (XSS), or browser-specific security flaws to execute malicious code through carefully crafted PDFs~\cite{pdf_nsfocous,SOCRadar2024}.

Machine learning (ML) is now extensively applied in various security contexts, including traffic detection, intrusion detection, vulnerability search, and other critical areas. In the face of PDF malware, numerous countermeasures have explored ML-based approaches~\cite{smutz2012malicious, smutz2016when, maiorca2015structural, vsrndic2013detection, vsrndic2016hidost, tong2019improving, chen2020training}. However, despite significant progress in PDF malware analysis, several critical challenges remain unresolved.

Firstly, one major challenge is the limited scope of existing feature analysis, which tends to be confined to surface-level inspection. Unlike the advanced feature analysis methods employed in binary code analysis—where researchers strive to extract semantic features from disassembled code and structural features from control flow graphs (CFGs)—PDF malware analysis often remains comparatively rudimentary. Current approaches typically involve computing specific keywords~\cite{smutz2012malicious} or analyzing structural paths~\cite{vsrndic2013detection}, which, while useful, fall short of the sophistication needed for gaining deeper insights. 
While the structure of PDF files differs from that of executable file formats, it presents its own set of unique complexities. 
Therefore, there is a pressing need for more advanced analysis techniques to effectively address the nuances of PDF malware.

Secondly, the superficial nature of current feature analysis leaves existing PDF malware classifiers highly susceptible to adversarial attacks~\cite{vsrndic2014practical, xu2016automatically, dang2017evading, maiorca2019towards}. 
While some studies~\cite{tong2019improving, chen2020training} have attempted to enhance the adversarial robustness of ML-based classifiers through specialized techniques like adversarial training, they have not substantially advanced feature engineering. Instead, these efforts continue to rely on the simplistic features extracted in earlier work~~\cite{smutz2012malicious, vsrndic2013detection}.
Research~\cite{maiorca2013looking, vsrndic2014practical, xu2016automatically} has consistently demonstrated that these features are vulnerable to adversarial manipulation. Additionally, retraining PDF malware classifiers using adversarial samples has significantly compromised their usability,  leading to false positive rates (FPR) as high as 85\%~\cite{AdversarialEvans}.

Lastly, the feature extraction process of existing machine learning-based PDF malware classifiers~\cite{Laskov2011Static,liu2014detecting,vsrndic2016hidost,chen2020training} is heavily dependent on the parsing capabilities of PDF parsers~\cite{carmony2016extract,anantharaman2023polydoc}. PDF malware that exploits vulnerabilities in PDF readers often results in so-called "bad PDFs." Attackers craft the original bytes of these PDFs to exploit specific vulnerabilities, which frequently results in malformed file formats~\cite{anantharaman2023polydoc}.
Such malformed PDFs can challenge existing parsers, rendering them unable to correctly process these files, even though the malicious payload can still be successfully executed. Carmony et al.~\cite{carmony2016extract} tested multiple parsers on PDF malware datasets and found that each parser failed to correctly process hundreds of samples. Consequently, PDF malware classifiers relying on these parsers cannot extract features from these samples, making it impossible to determine their maliciousness. This limitation is unacceptable in practical applications, where comprehensive detection and analysis of malware are crucial.

To overcome the above first issue, we design the first intermediate language framework, termed \textit{PDFObj IR}, to convert PDFs into a CFG-like structure. We observed that in PDFs, individual objects operate similarly to basic blocks in traditional program analysis, with reference relationships linking these objects. 
This similarity enables the construction of a graph structure akin to a CFG for effective analysis. Leveraging this analogy, we developed PDFObj IR, which converts each PDF object into a form similar to assembly language, describing each key-value pair in the object while preserving the inter-object reference relationships. Building on this framework, we constructed an Object Reference Graph (ORG), which allows for binary-like analysis of PDFs.

To address the second issue, we developed the PDFObj IR representation learning method, \textit{PDFObj2Vec}, a novel PDF feature engineering approach. 
We designed three representation learning schemes for this PDFObj2Vec, based on Word2Vec~\cite{mikolov2013distributed}, PV-DM~\cite{le2014distributed}, and BERT~\cite{devlin2019bert}. Additionally, we supported PDFObj2Vec with general text embedding models such as CodeT5~\cite{wang-etal-2021-codet5} and text-embedding-3~\cite{openai_embedding}, to directly obtain embeddings of PDFObj IR at the ORG node level.
We then designed a Graph Isomorphism Network (GIN) to extract structural features at the graph level of the ORG for PDF malware classification. 

Graph structures typically exhibit stronger adversarial robustness, making it challenging for attackers to disguise their behavior within such structures~\cite{akoglu2015graph}. Therefore, our feature engineering approach, which combines semantic and structural features, demonstrates robust performance. This robustness is evidenced by our experimental results, which show strong resilience against various adversarial attacks.
While language models have been extensively researched in the context of binary code analysis~\cite{massarelli2019safe,zuo2018neural,chua2017neural,ding2019asm2vec,li2021palmtree}, their application to PDF malware analysis has been limited. Our research bridges this gap by applying popular language models, including large language models, to PDF malware analysis. Previous features did not integrate well with these language models, but PDFObj IR demonstrates excellent compatibility, improving the performance of PDF malware analysis tasks.

To address the third issue, we developed a new PDF parser tool for extracting and converting PDFObj IR, called \textit{Poir}. Poir is immune to bad format issues affecting conventional parsers. 
By analyzing various types of bad-format PDF files, we identified three main types of errors that cause PDF parsers to fail. Poir automatically detects and fixes these errors, ensuring smooth feature extraction.

We applied ORG and PDFObj2Vec to the task of PDF malware classification and implement a GIN-based classifier that achieves well consistent performance on both the baseline and extended datasets. Our most robust classifier attains an accuracy of 99.93\% on the baseline dataset and 96.62\% on the extended dataset.
We also conducted extensive comparative and ablation experiments. 
The results indicate the effectiveness of PDFObj IR in PDF malware analysis, as our classifiers achieved a 2.2\% to 8.9\% accuracy improvement on the extended dataset compared to classifiers without PDFObj IR. 
Furthermore, even when faced with the most powerful realizable adversarial sample attacks, our classifier maintained 100\% adversarial robustness with a remarkably low FPR of only 0.07\%. This performance is significantly more efficient compared to state-of-the-art PDF malware classifiers~\cite{chen2020training,tong2019improving} with comparable adversarial robustness, whose FPR is 71 times higher than ours.
In a nutshell, we make the following key contributions:

\begin{itemize}
   \item We designed the PDFObj IR framework, which, to the best of our knowledge, is the first intermediate representation used for PDF analysis. To facilitate IR conversion, we developed a new PDF parser, Poir. This parser is capable of correctly handling malformed PDFs and automatically completing missing content.
   \item We developed PDFObj2Vec, a method that utilizes language models to learn representations of PDF objects. This approach was applied to PDF analysis, with a particular focus on evaluating its performance in PDF malware classification tasks.
   \item Leveraging the Object Reference Graph and PDFObj2Vec, we implemented a Graph Isomorphism Network for PDF malware classification. Experimental results demonstrate that our approach achieves high accuracy and strong adversarial robustness, all while maintaining an exceptionally low false positive rates.
\end{itemize}

\noindent \textbf{Open Source }
We release a prototype of PDFObj2Vec and evaluation datasets to facilitate reproduction, as all are found at \href{https://zenodo.org/records/15532394}{\underline{Zenodo}}.

\section{Background, Motivation and Related Work}

\subsection{PDF Basics}
\label{pdf_struct}

\begin{figure*}
    \centering
    \includegraphics[width=1\linewidth]{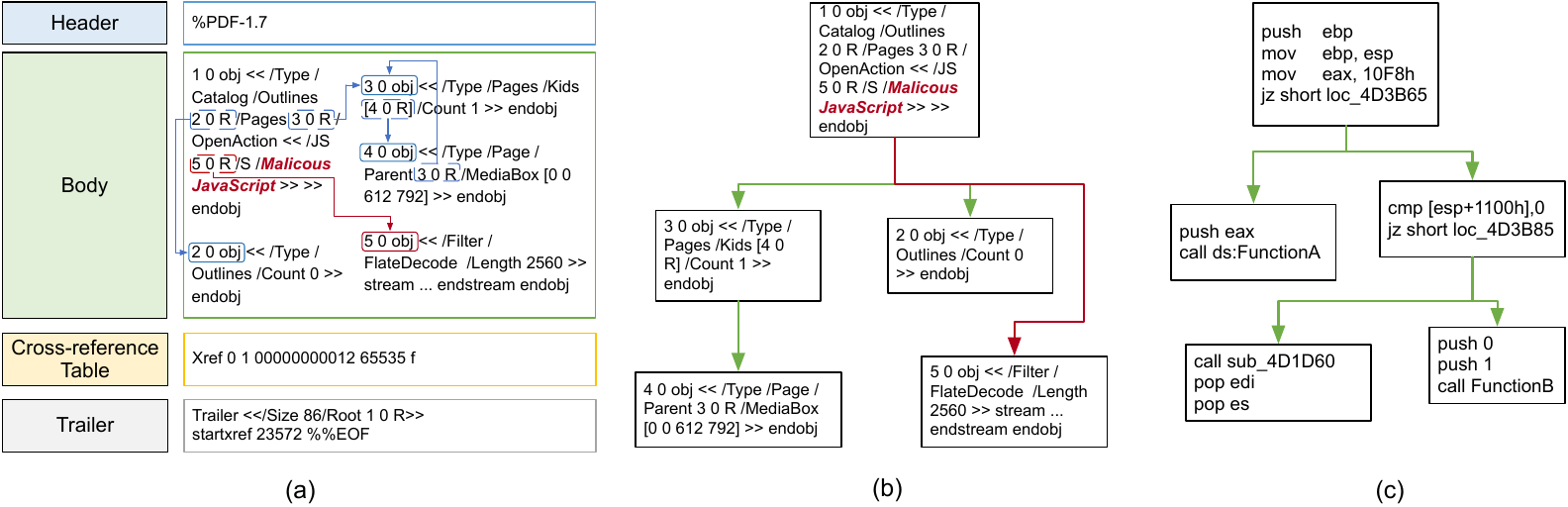}
    \vspace{-5mm}
    \caption{An example of the PDF structure, PDF object graph, and control flow graph: (a) displays the basic format of a PDF; (b) depicts the references between objects in PDF; (c) showcases a piece of CFG of a binary file.}
    \label{fig:pdf_background}
    \vspace{-2mm}
\end{figure*}

\noindent \textbf{PDF Structure} PDF is one of the most commonly used document formats on the web~\cite{pdfa_popularity}, with widespread applications in both personal and business contexts. Figure~\ref{fig:pdf_background}(a) illustrates a typical structure of a PDF file, which consists of the following four parts~\cite{pdf_standard}:
\begin{itemize}
    \denseitems
    \item \textbf{Header }This is the first line of a PDF file, specifying the version of the PDF specification used for the document.
    \item \textbf{Body }The body of a PDF file comprises various types of objects, forming a collection of objects. 
    The core component of a PDF file is the collection of these objects, also known as COS (Carousel Object System)~\cite{lowagie2010itext} objects.
    \item \textbf{Cross-reference Table }This table contains references to all objects within the document, listing the byte offsets of each object within the file's body. 
    \item \textbf{Trailer }
    The trailer enables quick identification of the cross-reference table's location, thus facilitating precise object location. The last line of the file only contains the file's end symbol: \texttt{\%\%EOF}.
\end{itemize}

The Body is the key part of a PDF, containing the primary data in the document. It comprises a series of objects, with each object enclosed by the $\ll$ and $\gg$, demarcated by the \texttt{obj} and \texttt{endobj}. Conceptually, it can be viewed as an object graph, where each object performs specific operations (e.g., displaying text, rendering images, executing code, etc.)~\cite{maiorca2019towards}.
Each object is composed of a series of key-value pairs, which can be represented in the form of a dictionary. 
For instance, in Figure~\ref{fig:pdf_background}(a), the first object begins with \texttt{1 0 obj} and ends with \texttt{endobj}. The content of this object is enclosed within $\ll$ $\gg$ and contains four key-value pairs. 
The keys in the object are the name type, and the values can be any type. 
For example, The first key in \texttt{1 0 obj} is \texttt{/Type}, which is a name type with the value of \texttt{/Catalog}, also a name type. 
The second key, \texttt{/Outlines}, has a value that is an indirect reference type, where \texttt{R} signifies an indirect reference.
In a PDF object, values encompass five categories of basic types, as shown in Table~\ref{tab:valuetype} of Appendix~\ref{basic_desc}. 
Moreover, values can be composite types, such as arrays and dictionaries, with the basic elements of arrays and dictionaries being the aforementioned basic objects, compression parameters, and other information.

\vspace*{2pt}
\noindent \textbf{PDF-based Attack }PDF-based attack is a type of document-based attack where threat actors exploit PDFs as carriers for malicious activities. These attacks leverage the functionalities of PDF files or vulnerabilities in PDF readers to execute malicious code. PDF malware refers to PDF carriers with malicious functionalities.

The body of Figure~\ref{fig:pdf_background}(a) is an example of a PDF malware that exploits JavaScript to execute malicious activities.
In Figure~\ref{fig:pdf_background}(a), we illustrate five objects, among which the first object's \texttt{/OpenAction}'s value encompasses information about malicious payloads. The value of \texttt{/OpenAction} is a dictionary composite type containing two keys: \texttt{/JS} and \texttt{/S}. 
Here, the value of \texttt{/JS} is \texttt{5 0 R}, indirectly reference to \texttt{5 0 obj}, and the value of \texttt{/S} is \texttt{/JavaScript}. 
The value indicates the presence of JavaScript within this PDF malware, with the scripts located at \texttt{5 0 obj}. The object of \texttt{5 0 obj} is typically a stream object, which stores the malicious JavaScript of this PDF malware.

In this example of PDF malware, fields related to JavaScript semantics are key indicators of malicious characteristics. Additionally, the JavaScript data are stored in \texttt{5 0 R}, indicating a reference relationship between \texttt{1 0 obj} and \texttt{5 0 obj}. 
In Figure~\ref{fig:pdf_background}(a), we use red arrows to indicate this reference relationship and blue arrows to mark the references between other objects.
Figure~\ref{fig:pdf_background}(b) more intuitively demonstrates the structural relationship between these objects in Figure~\ref{fig:pdf_background}(a).
Fields related to JavaScript semantics and the reference structure collectively constitute the malicious characteristics. 
However, semantic features alone are insufficient to determine whether a PDF is malicious, as JavaScript functionality is common in PDFs and can be used in benign samples. Therefore, it is also necessary to consider the structural relationships between objects. In Figure~\ref{fig:pdf_background}(a), the malicious JavaScript script data is stored in the stream object \texttt{5 0 obj}, which is used for complex malicious purposes. In contrast, simple JavaScript scripts in benign PDFs may appear as literal strings in \texttt{1 0 obj}, with no indirect reference to a stream object. These differences in structural reference relationships help further distinguish malicious PDFs from benign ones.

\subsection{Motivation and Insight}

There are rich features used for binary malware analysis, such as API call sequences, control flow graphs (CFGs), data flow graphs, and disassembly instructions. In contrast, analytical features for PDF malware analysis are relatively limited. Current methods often rely on customized statistical features of keywords or objects, or on binary features derived from hierarchical structural paths.

This disparity raises an important question: can we apply the methods used in binary malware analysis to PDF malware analysis? To explore this, we reexamined the structure of PDFs, which are primarily composed of a series of objects. These objects have complex reference relationships, as indicated by the blue arrows in Figure\ref{fig:pdf_background}(a). Essentially, this forms a directed graph connected by different objects. If we consider each object as a basic block of a PDF and the reference relationships between objects as control flow relationships, the resulting graph, as shown in Figure~\ref{fig:pdf_background}(b), would resemble a control flow graph used in binary analysis, as shown in Figure~\ref{fig:pdf_background}(c). This similarity suggests that leveraging methods from binary analysis to analyze PDFs is feasible. However, a critical question remains: how do we represent a basic block in the context of PDF analysis?

In binary analysis, the disassembled code is often scrutinized. For instance, when analyzing a sample in IDA Pro~\cite{idapro}, the software typically presents a CFG of the entry point function after loading the sample. Each basic block corresponds to a disassembled code block, representing the smallest unit of code, as illustrated in Figure~\ref{fig:pdf_background}(c). This raises the question: can we transform the content of a PDF object into a form similar to disassembled code or intermediate code?
To explore this, we examined the structure of PDF objects. These objects consist of key-value pairs with fixed data formats and types. By defining a fixed format to describe each object in a manner akin to natural or programming languages, we can convert the object into an intermediate representation. We can then designate the values containing reference relationships as jump instructions. This approach allows us to construct a graph structure that is closely analogous to a CFG, which we refer to as an Object Reference Graph (ORG).
Given that we propose converting objects into an intermediate representation, we can leverage powerful language models to process this representation. In the field of program analysis, there has been considerable research using language models to represent disassembled code. We will elaborate on our design philosophy and processing approach in the following sections.

\vspace*{2pt}
\noindent \textbf{Key Insight }Our key insight is that by treating each object in a PDF as a basic block and converting its content into an intermediate language, we can construct a graph analogous to a CFG. This graph encapsulates the reference relationships among different objects, with each node representing the semantics of an object. Thus, this approach enables us to analyze PDFs both at the semantic level of individual nodes and at the structural level of the graph.

\subsection{PDF Malware Analysis}
Currently, research on PDF malware analysis mainly builds upon two prior works: PDFrate~\cite{smutz2012malicious} and Hidost~\cite{vsrndic2013detection}. 
PDFrate utilizes content-based features, extracting specific keyword positions and counts from metadata and content within PDF files. It manually defines 202 features, which are extracted using regular expressions. While these features are more general and not affected by the parsing capabilities of parsers, they remain at a surface level, not delving into the deeper structure of PDFs. Moreover, defining these 202 features requires extensive expert knowledge, and the reliability of these features is not always guaranteed~\cite{vsrndic2014practical,tong2019improving}.
Hidost, on the other hand, employs structure-based features by extracting object structural paths from PDFs and using binary counts of these paths as features. It leverages Poppler~\cite{Poppler55} to extract hierarchical structural paths and selects $6,087$ paths from a corpus of 9 million as features. Despite the authors' claims that hierarchical path features are robust~\cite{vsrndic2013detection, vsrndic2016hidost}, selecting only a portion of the paths from the corpus may result in a lack of path semantics. Xu et al.'s study~\cite{xu2016automatically} indicates that such features remain vulnerable to adversarial attacks. Furthermore, Hidost relies on Poppler for PDF parsing, which makes feature extraction more susceptible to the parsing capabilities of the parser.

Due to the susceptibility of PDFrate and Hidost to adversarial attacks, researchers have sought to enhance the adversarial robustness of PDF malware classifiers based on their features. Tong et al.~\cite{tong2019improving} proposed a PDF malware classifier that leverages conserved feature training, focusing on features derived from PDFrate and Hidost. They identified features closely related to malicious functionalities in PDF execution as conserved features through expert experience. By employing iterative adversarial training, they improved the classifier's adversarial robustness. However, while Tong et al.'s method enhances robustness against adversarial attacks, it compromises classification performance on regular samples, resulting in a higher FPR of 4.96\%.

Additionally, Chen et al.~\cite{chen2020training} proposed a robust training approach based on robust properties targeting the features of Hidost. They defined five categories of robust properties and used symbolic interval analysis to train combinations of different robust properties, resulting in a classifier with adversarial robustness. However, Chen et al.'s method, while achieving adversarial robustness, also sacrifices the classification performance on regular samples, increasing the FPR by 1.78\%. Moreover, when facing state-of-the-art unbounded adversarial sample attacks~\cite{maiorca2013looking}, it only achieved a 50.8\% adversarial success rate. Both Tong et al. and Chen et al. attempted to enhance the classifier's adversarial robustness based on PDFrate and Hidost features, addressing the training issues of the classifier. 
However, they did not resolve the fundamental problem of insufficiently robust features and the limitations imposed by the parser dependencies in classifiers based on Hidost.

\subsection{Learning-based Embedding}

Drawing inspiration from representation learning in binary analysis~\cite{massarelli2019safe,zuo2018neural,chua2017neural,ding2019asm2vec}, we aim to develop semantic representations for PDFObj IR nodes in the ORG and apply them to downstream tasks such as PDF malware detection. Notably, this type of representation learning remains unexplored in PDF analysis.
In binary analysis, approaches like Word2Vec~\cite{mikolov2013distributed} have been used to learn instruction-level representations by treating each instruction as a word and each function as a document~\cite{massarelli2019safe,zuo2018neural}. Asm2Vec~\cite{ding2019asm2vec} extends this by representing assembly instructions as opcodes and operands, using the PV-DM model~\cite{le2014distributed} to learn embeddings. 
Similarly, PalmTree~\cite{li2021palmtree} treats assembly instructions as sentences, decomposing them into tokens (e.g., opcodes, registers, immediate values) and employing BERT~\cite{devlin2019bert} to capture control flow and data dependencies.

In this paper, we applied Word2Vec, PV-DM, and BERT to learn representations for PDFObj IR, generating embeddings for downstream tasks. Additionally, we used general embedding models without pre-training to directly derive embeddings for PDFObj IR. The evaluation of these embedding models in PDF malware classification is detailed in $\S$\ref{evaluation}.

\section{Overview}

\begin{figure}[]
    \centering
    \includegraphics[width=1\linewidth]{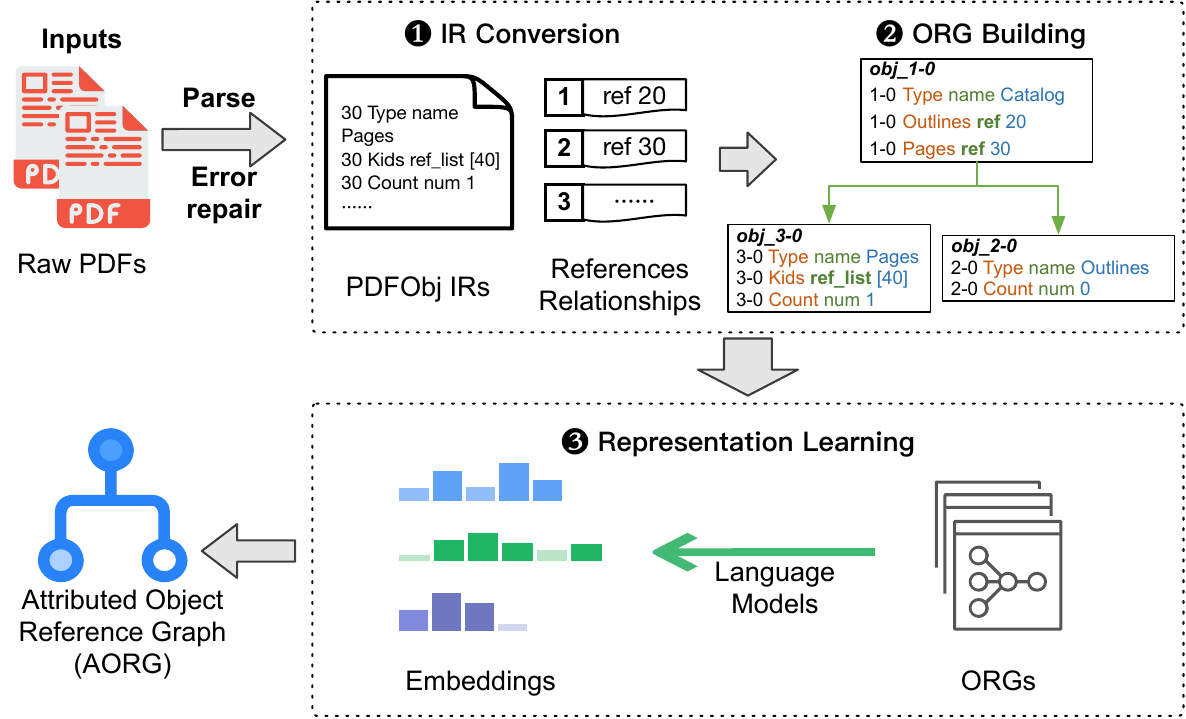}
   \vspace{-3mm}
    \caption{Overall workflow of PDFObj2Vec.}
    \label{fig:overview}
    \vspace{-5mm}
\end{figure}

The overall workflow of PDFObj2Vec is depicted in Figure~\ref{fig:overview}. This framework primarily consists of two parts: 1) PDFObj intermediate representation (IR) conversion and 2) representation learning. 

\vspace*{2pt}
\noindent \textbf{PDFObj Intermediate Representation (IR) Conversion }For the conversion of PDFObj to IR, PDFObj2Vec starts by taking a raw PDF file as input, then parses it, correcting formatting errors to ensure the integrity of the extracted IR. The parsed content is then converted into IR format (\circled{1} in Figure~\ref{fig:overview}). Following this, based on the reference relationships in the IR, an Object Reference Graph (ORG) is constructed (\circled{2}). In the ORG, each node represents an object, with the node's content being the IRs of that object. 

\vspace*{2pt}
\noindent \textbf{Representation Learning }
In the process of representation learning, we embed the nodes of the ORG (\circled{3}). To achieve this, We pre-trained Word2Vec, PV-DM, and BERT models specifically for PDFObj IR to obtain node embeddings. Additionally, we integrated general embedding models such as the standard BERT, CodeT5, and text-embedding-3 to directly obtain node embeddings.
Once the node embeddings are obtained, we can generate the Attributed Object Reference Graph (AORG). Then, we can perform downstream tasks such as PDF malware classification.

\section{PDFObj IR Conversion}
\label{pdfobjir_design}

We designed the first intermediate representation (IR) framework for PDFs, aimed at enhancing the analysis and understanding of PDFs by enriching the semantics of the objects of PDFs. In this framework, each object in a PDF is converted into multiple fixed-length IRs to represent the corresponding object. 
In this section, we first define the fields and conversion rules of the IR, and then we introduce the new parser we developed for this purpose, named \textit{Poir}.

\subsection{Field Definitions}
Based on the structure of objects, we define the four fields of PDFObj IR: Index, Attribute, VType, and Value, as described in the following:

\begin{itemize}
    \denseitems
    \item \textbf{Index.} This field indicates the index of the current IR within the object. Its value is uniquely determined by the combination of the object's number and version, calculated as $<number-version>$. 
    For instance, all IRs within the object \texttt{3 0 obj} have an Index of 3-0, and within \texttt{8 2 obj}, all IRs have an Index of 8-2.
    \item \textbf{Attribute.} This field represents the attributes of the IR, corresponding to the keys in the associated object. It is of the Name type, such as \texttt{/Type} and \texttt{/Pages}.
    \item \textbf{VType.} This field represents the type of the Value. We have defined a total of 6 atomic types, 1 stream type, 2 composite types, and 6 derived types, as shown in Table~\ref{tab:IRtype_desc}.
    \item \textbf{Value.} This field represents the values associated with the Attribute. The type of Value is VType, and we defined 15 VTypes, as shown in Table~\ref{tab:IRtype_desc}.
\end{itemize}


Among these VTypes in Table~\ref{tab:IRtype_desc}, \texttt{list} and \texttt{dict} are basic composite types, and their values may consist of a mix of multiple atomic types and basic composite types. 
Elements in \texttt{list} are typically of a single type in most cases, but occasionally, mixed-type values occur. Based on this phenomenon, we defined six derived types based on \texttt{list}, as shown in Table~\ref{tab:IRtype_desc}. Please note that we have not defined \texttt{null\_list} as it does not exist. 
Since a \texttt{dict} consists of a series of key-value pairs of various types, it is inherently a structure with mixed types. Therefore, there is no need to design derived types based on \texttt{dict}.
We have defined the format and basic fields of PDFObj IR, with 15 types for the VType field. Thus, each object can be represented using $n$ IR entries. In the next subsection, we  
provide a detailed description of how we convert a PDF into a series of IR entries.

\begin{table}[]
  \centering
  \caption{Basic type description.}
  \vspace{-3mm}
  \begin{small}
  \label{tab:IRtype_desc}
  \begin{tabular}{cc}
    \toprule
    Type & VType Mark \\ 
    \midrule
    Atomic & \texttt{num, str, name, ref, bool, null}\\
    Stream & \texttt{stream}\\
    Composite & \texttt{list, dict}\\
    \multirow{2}*{Derived} &  \texttt{num\_list, str\_list, name\_list}\\
                        &\texttt{ref\_list, bool\_list,  mix\_list}\\

  \bottomrule
\end{tabular}
\end{small}
\vspace{-2mm}
\end{table}

\subsection{Conversion}

\begin{figure} []
    \centering
    \includegraphics[width=0.75\linewidth]{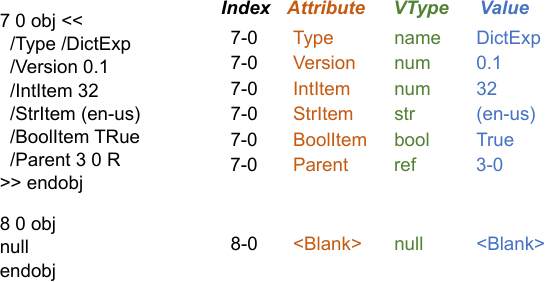}
     \vspace{-2mm}
    \caption{Basic conversion paradigm. Please note that this example is designed to illustrate the basic conversion principle and may not necessarily represent data found in actual PDFs.}
    \label{fig:conversion_p1}
    \vspace{-5mm}
\end{figure}

Following the definition of fields and formats of PDFObj IR, we initiate the conversion of each key-value pair within an object into $n$ IR entries, where $n \geq 1$. The conversion form varies according to the VType. 
We define the atomic IR as follows:

\textbf{Definition 1:} \textit{An atomic IR is an IR with an atomic type or a value of the basic object, representing the most basic expression of PDFObj IR that cannot be further decomposed.}

We first use the atomic IR conversion as an example to illustrate the basic conversion principle. Its conversion is the most direct and fundamental. We tend to convert each PDFObj IR into an atomic form, ensuring that each IR maintains the same structural form. 

\vspace*{2pt}
\noindent \textbf{Basic Conversion Paradigm (P1) }For basic object types such as numeric, string, name, boolean, null, and ID object, as mentioned in $\S$\ref{pdf_struct}, the VType is assigned as \texttt{num}, \texttt{str}, \texttt{name}, \texttt{bool}, and \texttt{null}, respectively, with the Value maintaining its original form. For an indirectly referenced object, its type is assigned as \texttt{ref}, and the value is the index value of the referenced object. Figure~\ref{fig:conversion_p1} illustrates a basic conversion paradigm, which serves as the foundational transformation. All subsequent complex conversions are built upon this paradigm. The conversion process varies depending on whether the object is a stream, dictionary, single-element array, or mixed-element array, as will be discussed in detail later.

\vspace*{2pt}
\noindent \textbf{Stream Object Conversion Paradigm (P2) } A stream object consists of two parts. The first part utilizes a dictionary to store basic information about the stream, such as encoding method, stream length, etc. The second part is the byte sequence data of the stream. When the object is a stream, the conversion process begins with outputting the following line:
\begin{verbatim}
    <Index>, <Blank>, stream, <Blank>
\end{verbatim}
The IR above declares that it is a stream object. Subsequently, the conversion of the first part of the dictionary content follows \textit{P1}. The byte sequence data are stored in an additional data file, named according to its Index.

\vspace*{2pt}
\noindent \textbf{Array Object Conversion Paradigm (P3) } As illustrated in Figure~\ref{fig:pdf_background}(a), the value of \texttt{/MediaBox} in \texttt{4 0 obj} is an array of single elements, all numeric. For this, we use the following representation:
\begin{verbatim}
    4-0, /MediaBox, num_list, [0,0,612,792]
\end{verbatim}
The conversions of other single-element arrays follow the same way. Arrays can also contain a mix of types, such as \texttt{/Names [(Notice) 14 9 R]}, featuring \texttt{str} and \texttt{ref} types. We represent this pair as the following IR:
\begin{verbatim}
    4-0, /Names, mix_list, [(Notice),149]
\end{verbatim}

\vspace*{2pt}
\noindent \textbf{Dictionary Object Conversion Paradigm (P4) } Take for instance the \texttt{/OpenAction} value in \texttt{1 0 obj} from Figure~\ref{fig:pdf_background}(a), which is a dictionary. We begin with an IR entry:
\begin{verbatim}
    1-0, /OpenAction, dict, <Blank>
\end{verbatim}
The IR above indicates that the value of \texttt{/OpenAction} is a dictionary type. When converting such dictionaries, we prepend the key of this \texttt{dict} (\texttt{/OpenAction} in this case), to the new Attribution, resulting in:
\begin{verbatim}
    1-0, /OpenAction/JS, ref, 5-0
    1-0, /OpenAction/S, name, /JavaScript
\end{verbatim}
Please note that in the conversion example above, a \texttt{dict} may contain another \texttt{dict}, resulting in multi-level nested dictionaries. We employ a recursive algorithm to resolve such nested dictionaries, ensuring that the IR can record the path of keys within nested dictionaries. 

\begin{figure}[t]
    \centering
    \includegraphics[width=0.9\linewidth]{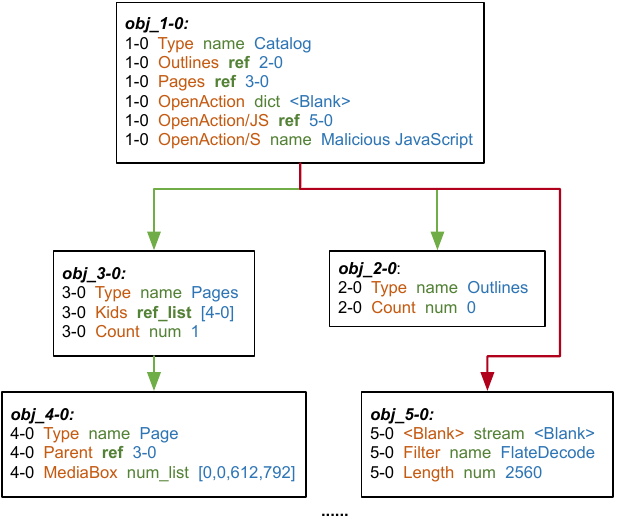}
    \vspace{-3mm}
    \caption{Object reference graph with PDFObj IR.}
    \label{fig:org3}
    \vspace{-5mm}
\end{figure}

So far, we have achieved a comprehensive conversion of the PDF's core content. We can convert the representation from Figure~\ref{fig:pdf_background}(b) to that of Figure~\ref{fig:org3}. Subsequently, we can analyze the PDF from both semantic and structural perspectives.


\subsection{PDFObj IR Parsing}

After finalizing the design and conversion paradigm of PDFObj IR, we developed a parsing tool named Poir. This tool converts PDF files into an intermediate representation format and is uniquely designed to handle malformed PDFs by performing necessary repairs.
PDF malware, particularly those exploiting vulnerabilities, often fails to maintain a valid PDF format, causing traditional parsers to malfunction. To address this, we analyzed malformed PDF malware and identified three common types of errors in the PDF body. Poir incorporates specific processes to handle these errors, ensuring the integrity of the IR is preserved as much as possible.

\label{error_process}

\begin{figure}[t]
    \centering
    \includegraphics[width=0.85\linewidth]{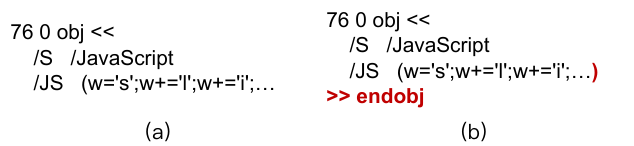}
    \vspace{-3mm}
    \caption{An example of string overflow (a) and the completion method (b).}
    \label{fig:error1}
    \vspace{-3mm}
\end{figure}

\vspace*{2pt}
\noindent \textbf{E1: String Overflow } String content overflow is the most common exception, and it occurs frequently in malicious PDFs. Figure~\ref{fig:error1}(a) illustrates a typical situation of string overflow where the excessive length of a string causes the omission of crucial keywords such as the right parenthesis and \texttt{endobj}. The cause of string content overflow may be related to the content, often malicious code. In such cases, we automatically fill in the missing structure and keywords, appending ``)'' at the end. If the overflow results in the absence of ``$\ll$'' and \texttt{endobj}, we supplement them as well, as shown in Figure~\ref{fig:error1}(b).

\begin{figure}[]
    \centering
    \includegraphics[width=0.6\linewidth]{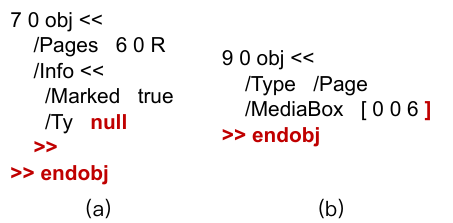}
    \vspace{-3mm}
    \caption{An example of incomplete key-value pairs and the completion method.}
    \label{fig:error2}
    \vspace{-5mm}
\end{figure}

\vspace*{2pt}
\noindent \textbf{E2: Mising \texttt{obj} } 
In the event of this error, an illegal indirect reference occurs. For instance, in the case of \texttt{/Metadata 9 0 R}, it references to a non-existent \texttt{obj}. We can not deduce the specific content of \texttt{9 0 obj}. Therefore, we introduce a new \texttt{obj} with the number and version set to 9 and 0, respectively, and the content is set to null.

\vspace*{2pt}
\noindent \textbf{E3: Incomplete Key-Value Pairs } 
Our analysis has identified cases where either the key or the value is missing. The corresponding value is also missing when the key is absent. To address this, we fill the original key-value position with null as the missing value, as illustrated in Figure~\ref{fig:error2}(a). In cases where the value part is incomplete, such as an array with only the front portion, as depicted in Figure~\ref{fig:error2}(b), we append `]' at the end to complete its structure. For incompletely nested dictionaries, we first supplement the missing parts based on the aforementioned principles and subsequently complete the structure of the dictionary. Additionally, if the omission of key-value pairs results in the absence of ``$\gg$'' and \texttt{endobj} in the object, we also rectify these omissions. 

In addition to these three main error handling measures, we have also listed other parsing errors and their corresponding handling strategies in Table~\ref{tab:appd_error_handling} of Appendix~\ref{other_error_process}.

\section{Design of PDFObj2Vec}
We have developed two modes for PDFObj2Vec: pre-trained mode and general mode. We first introduce the preprocessing and tokenization strategies for PDFObj IR. Subsequently, we outline the design of the pre-trained mode and general mode. Lastly, we discuss the ORG embedding and the classifier architecture based on PDFObj2Vec.

\subsection{Preprocessing and Tokenization}
In $\S$\ref{pdfobjir_design}, we fixed the length of PDFObj IR and have already standardized it, so we do not require additional special tokens for normalization.
To adapt PDFObj IR for pre-training, we need to perform tokenization. We employ the following tokenization strategy to mitigate the Out-Of-Vocabulary (OOV) issues caused by values: For each IR, we retain the Attribute and VType, and connect them with an underscore to form a single word. For example, in Figure~\ref{fig:conversion_p1}, \texttt{7-0 Type name DictExp} would be represented as \texttt{Type\_name}. Multiple IRs form an object, and multiple words form a sentence; therefore, we treat an object as a sentence. 
The contextual relationships between sentences are determined by the reference relationships between objects. We extract these relationships from the \texttt{ref}, \texttt{ref\_list}, and \texttt{mix\_list} (where \texttt{mix\_list} may include reference types) to generate the context of the sentences.

\begin{figure}[]
    \centering
    \includegraphics[width=1\linewidth]{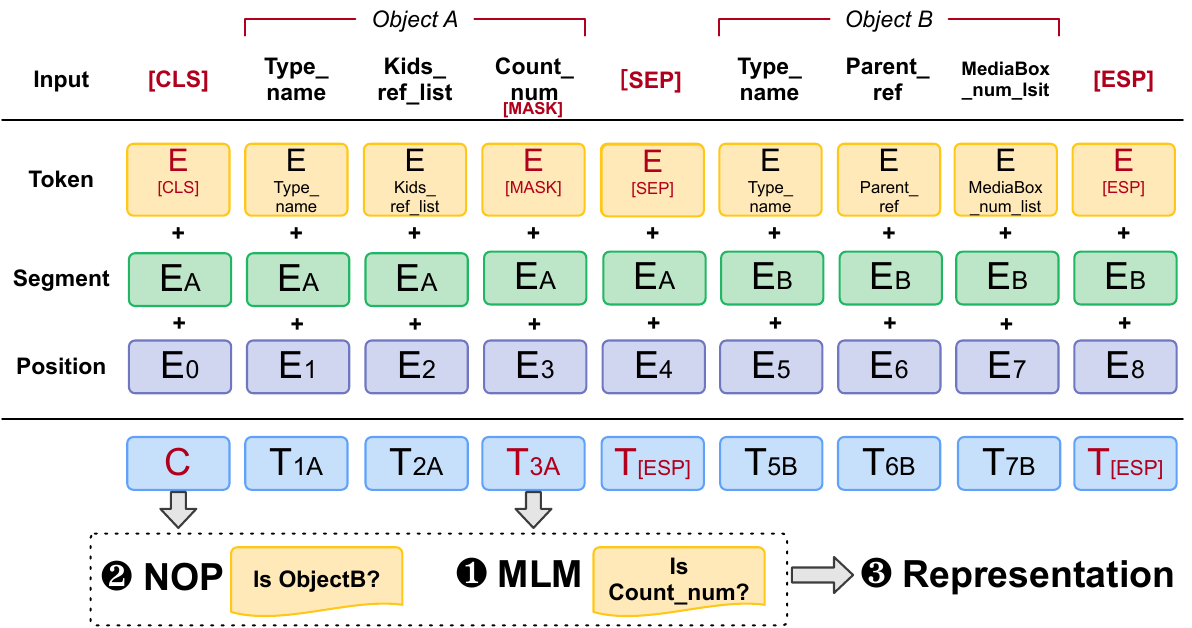}
    \vspace{-6mm}
    \caption{BERT input and training tasks.}
    \label{fig:bertinput}
    \vspace{-4mm}
\end{figure}

\subsection{Pre-trained Mode}

We devised three schemes for the pre-trained mode, namely Word2Vec, PV-DM, and BERT, with a particular emphasis on the BERT scheme.
Prior feature embedding methods, such as Hidost, rely on binary embeddings that generate sparse 0-1 vectors. These representations are inherently fragile and lack the rich semantic information required for robust malware detection. In this paper, we leverage learning-based embedding techniques from natural language processing. Trained via self-supervised learning tasks, these models convert the PDFObj IR into dense vector representations, thereby enhancing the classifier’s ability to detect PDF malware.

\vspace*{2pt}
\noindent \textbf{BERT-Based Scheme } 
After preprocessing and tokenization, we then input the sentences into the BERT model, as depicted in Figure~\ref{fig:bertinput}. The first token of this input is a special token, \texttt{[CLS]}, which signifies the start of the sequence. Following this, we use another token, \texttt{[SEP]}, to separate Object A from Object B. Additionally, we augment token embeddings with position embeddings and segment embeddings, and this combined vector is used as the input for the bidirectional transformer network, as shown in Figure~\ref{fig:bertinput}. Segment embeddings help BERT differentiate between the vector representations of the two sentences in the input, while position embeddings enable BERT to learn the sequential properties of the input. As for pre-training, we designed following two training tasks for BERT-based PDFObj2Vec: MLM (Masked Language Model, \circled{1} in Figure~\ref{fig:bertinput}) and NOP (Next Object Prediction, \circled{2}).

\vspace*{2pt}
\textit{\circled{1} Masked Language Model } 
To enable BERT to comprehend the internal structure of PDFObj IR, we first employed the Masked Language Model (MLM) training task. 
This task randomly masks tokens in the PDFObj IR text, forcing BERT to predict the masked content through bidirectional contextual inference. And it enables the model to learn deep semantic relationships between key-value pairs within objects.
We began by pre-training BERT-based PDFObj2Vec using the MLM, following masking strategies from previous studies~\cite{devlin2019bert, li2021palmtree}.
For the input IR sequences $Seq_{IR} = IR_1, IR_2, ..., IR_i$, in which $IR_i$ denotes a token, we randomly select 15\% of the tokens to be masked. 
Of these tokens to be masked, 80\% are replaced with the \texttt{[MASK]} token, 10\% are replaced with a random IR token, and the remaining 10\% are left unchanged. Subsequently, BERT's encoder learns to predict the masked tokens:
$$ P(\hat{IR_i} \mid Seq_{IR}) = \frac{exp(w_i \Theta (SeqIR)_i)}{\sum^{K}_{k=1}exp(w_k \Theta (SeqIR)_i))} $$
where \( \hat{IR}_i \) represents the prediction for \( IR_i \), \( \Theta(SeqIR)_i \) denotes the \( i \)-th vector from the last hidden layer of the transformer network \( \Theta \), \( w_i \) represents the weight, and \( K \) is the size of the vocabulary. 
The loss $\mathcal{L}_{MLM}$ for this task is the cross-entropy loss.
Given a PDFObj IR pair, Object A and Object B, we first add special tokens \texttt{[CLS]} and \texttt{[SEP]}, and then replace the token for \texttt{Count\_num} with a \texttt{[MASK]} token. Next, we input this modified PDFObj IR pair into the BERT model, which will then make predictions for the \texttt{[MASK]} token, as shown in Figure~\ref{fig:bertinput}.

\vspace*{2pt}
\textit{\circled{2} Next Object Prediction } 
To enable BERT to capture the reference relationships between objects, we designed a training task based on the Next Sentence Prediction called Next Object Prediction (NOP). 
This task treats inter-referenced object IRs as continuous sequences and trains BERT to determine the likelihood of object reference relationship. Through this process, BERT learns the logical structure of the ORG graph.
When constructing the input PDFObj IR pair for the NOP task, we select pairs with real reference relationships with a probability of 50\%.
Specifically, we input two objects: $obj_1$ and $obj_2$, starting with the \texttt{[CLS]} token and separated by a \texttt{[SEP]} token. 
This method trains the BERT-based PDFObj2Vec model to predict the probability that a reference relationship exists between two objects:
$$P(\hat{y} \mid obj_1, obj_2) = \frac{expS(y \mid obj_1, obj_2)}{\sum_{y \in (0,1)} exp S (y \mid obj_1, obj_2)}$$
where $\hat{y}$ denotes the reference relationship prediction of $obj_1$ and $obj_2$, $S()$ denotes the function of attention head in transformer. 
The loss $\mathcal{L}_{NOP}$ for this task is the cross-entropy loss.
We select the first output vector in Figure~\ref{fig:bertinput} to predict whether the two objects have a reference relationship.
In the case of Figure~\ref{fig:bertinput}, where there is a reference relationship between the two objects, the correct prediction would be 1; otherwise, it would be 0.
The loss function of BERT scheme is the combination of $\mathcal{L}_{NOP}$ and $ \mathcal{L}_{MLM}$.

\vspace*{2pt}
\textit{\circled{3} PDFObj IR Representation } 
Through the MLM and NOP pre-training tasks, BERT-based PDFObj2Vec learns the semantics of the IRs of objects and references relationships between these objects in ORG. Then, we can generate contextually enriched object embedding vectors. 
Specifically, an object IR beginning with a \texttt{[CLS]} token is inputted into PDFObj2Vec-Bert.
We then compute the pooling of the \texttt{[CLS]} token at the last hidden layer of PDFObj2Vec-Bert and use this pooling value as the semantic embedding vector for the object IR.
This approach allows for a nuanced representation that captures both the individual characteristics of the object and its relational context within the ORG.

\vspace*{2pt}
\noindent \textbf{Word2Vec and PV-DM-Based Schemes } 
The preprocessing and tokenization for Word2Vec and PV-DM are consistent with those for BERT. 
In Word2Vec, we utilized the CBOW model that trains Word2Vec by predicting the center word from a given context.
The Word2Vec scheme can only generate embedding of the word, so to obtain the embedding of an object in a PDF, we use TF-IDF weighted averaging.
The fundamental idea behind the PV-DM model is similar to CBOW, and it combines paragraph vectors and context word vectors to jointly predict the target word and train the model accordingly. 
During the PV-DM training process, each object is treated as a paragraph composed of multiple words. The representation of the paragraph vector is similar to that of Word2Vec, which adopts a TF-IDF weighted average of word vectors in the paragraph.
Detailed model structures and parameter specifications are provided in Appendix~\ref{word2vec_pvdm_details}.

Transforming PDFObj IR text into vector embeddings through learning-based methods is essential, as it yields a substantially more effective feature set. These embeddings capture the intrinsic properties of PDF objects as well as the contextual dependencies between objects, both of which are critical for accurate and robust malware classification. Without this transformation, the resulting features would be too fragile and simplistic to effectively detect sophisticated adversarial attacks.

\subsection{General Mode}
\label{general_emb}
We also integrated three general embedding schemes to obtain embeddings for PDFObj IR: the BERT Base~\cite{devlin2019bert,standard_bert}, CodeT5~\cite{wang-etal-2021-codet5}, and text-embedding-3~\cite{openai_embedding}. These general embedding models have been trained on extensive and diverse corpora. We downloaded the BERT Base and CodeT5 models to compute the embeddings for PDFObj IR. 
We also integrated the most popular conversational model, ChatGPT's embedding model, text-embedding-3, which is OpenAI's latest third-generation embedding model. We used the API provided by OpenAI to obtain IR embeddings directly. For preprocessing and tokenization, we adopted the same methods used in the previous pre-trained models.

\subsection{Graph Embedding and Classification}
\label{gnn_classify}

\begin{figure}
    \centering
    \includegraphics[width=1\linewidth]{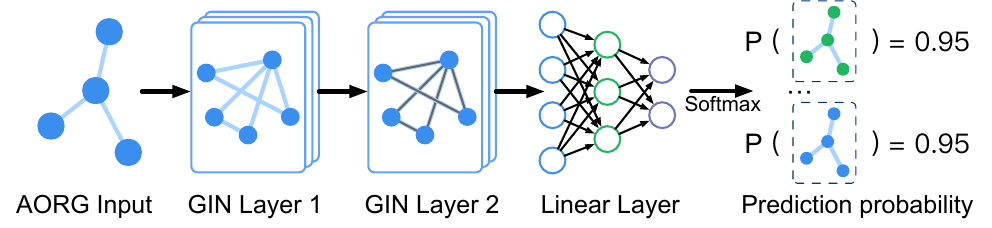}
     \vspace{-7mm}
    \caption{The network structure of our proposed GIN-based classifier.}
    \label{fig:ginstruct}
    \vspace{-5mm}
\end{figure}

After obtaining the semantic embeddings for all objects in the ORG, we transform it into a semantic Attributed Object Reference Graph (AORG) suitable for a graph neural network. We designed a Graph Isomorphism Network (GIN) to compute the graph representation of the AORG and classify PDF malware, as illustrated in Figure~\ref{fig:ginstruct}. The core concept of GIN is to aggregate the features of each node to capture the graph's topological structure.
We employ a Multi-Layer Perceptron (MLP) as the aggregation function, enabling GIN to optimally distinguish graph isomorphisms. This structure allows GIN to effectively learn and represent complex patterns within the data, facilitating accurate classification of the entire graph. Detailed parameters and technical specifications of the GIN classifier are provided in Appendix~\ref{GIN_Classifier}.

\section{Experimental Evaluation}
\label{evaluation}

We conducted a comprehensive set of experiments to evaluate the effectiveness of PDFObj IR and PDFObj2Vec. Our evaluation covered the following six aspects: 1) parsing capability of Poir; 2) evaluation of pre-trained PDFObj2Vec; 3) performance of pre-trained PDFObj2Vec in the PDF malware classification task; 4) performance of general PDFObj2Vec in the PDF malware classification task; 5) evaluation against adversarial attacks; 6) ablation studies.

\subsection{Experimental Setup}

\noindent \textbf{Dataset }We used the contagio~\cite{contagio13_online} dataset as our baseline dataset, which was commonly used in PDF malware analysis studies~\cite{chen2020training,tong2019improving,smutz2012malicious}. It has a balanced distribution of benign and malicious samples, comprising 9k benign samples and 11k malicious samples. 
Additionally, we collected an extended dataset for testing that includes 21k malicious samples from the CIC-PDFMal2022 dataset~\cite{issakhani2022pdf,CICdataset} and 24k benign samples gathered from the internet. These benign samples include a diverse array of PDF types, such as bills, test files, books, and interactive forms, obtained from sources like GitHub, the PDF Association~\cite{stressfulpdf}, and the gov PDF dataset~\cite{dotgovpdf}. 
We used MD5 checksums to ensure that there was no overlap between the baseline and extended datasets. And we also used timestamps to confirm that samples in the extended dataset were collected after those in the baseline dataset.

\vspace*{2pt}
\noindent \textbf{Baseline PDF Parsers }To compare the parsing completeness with Poir, we select six popular PDF parsing tools, namely pdfrw~\cite{pdfrw}, Poppler~\cite{Poppler55}, pdfminer~\cite{pdfminer}, MuPDF~\cite{mupdf}, borb~\cite{borb}, and QPDF~\cite{QPDF}, as baseline parsers. 
Among these six baseline parsers, pdfrw, pdfminer, and borb are written purely in Python, while Poppler and QPDF are written in C++, and MuPDF is written in C.

\vspace*{2pt}
\noindent \textbf{PDFObj2Vec Configurations }
We implemented two modes of PDFObj2Vec: pre-trained and general. For the first mode, we pre-trained PDFObj2Vec schemes based on Word2Vec, PV-DM, and BERT, setting the embedding dimension of each model to 512. The specific training hyperparameters for these three pre-trained models are provided in Appendix~\ref{Settings_PDFObj2Vec}. For the general mode, we applied three off-the-shelf general embedding models: BERT Base~\cite{devlin2019bert,standard_bert}, CodeT5~\cite{wang-etal-2021-codet5}, and text-embedding-3~\cite{openai_embedding}.
BERT Base is a general embedding model for natural language, CodeT5 is a pre-trained embedding model for programming languages based on the T5 architecture, and text-embedding-3 is OpenAI's latest third-generation embedding model. 
For these three schemes, we used their default embedding dimensions of 768, 256, and 1536, respectively.

\vspace*{2pt}
\noindent \textbf{PDF Malware Classifiers Configurations}
We reproduced two state-of-the-art PDF robust malware classifiers~\cite{chen2020training,tong2019improving}, which demonstrate excellent adversarial robustness. One is robustly trained classifier (RTC). Chen et al.~\cite{chen2020training} used symbolic interval analysis to robustly retrain a deep neural network based on the robustness attributes defined by the structural paths. The other is conserved features trained classifier (CFTC). Tong et al.~\cite{tong2019improving} adversarially retrained a support vector machine classifier based on the conservative features defined by the structural paths.
Both RTC and CFTC's reproduction details and hypermeters are provided in Appendix~\ref{baselines_classifiers_details}. 
We followed the training setup on the baseline contagio dataset, as used by RTC and CFTC. Our implementation of the GIN-based classifier was also trained on this baseline dataset with hyperparameter details provided in Appendix~\ref{GIN_Classifier}. 
All experiments were conducted on a machine with an Intel Core i7-12700K, NVIDIA GeForce GTX 3090, and 64 GB RAM running Ubuntu 20.04.

\subsection{Parser Performance}
We selected the latest versions of six baseline parsers, with the exception of pdfrw, which was last updated in 2018. Parsing tests were conducted on both the baseline dataset and an extended dataset containing malicious samples with malformed formats that do not impact their malicious functionality. The experimental results are presented in Table~\ref{tab:parse_results}. 
An error refers to a failure that occurs when the parser encounters a malformed PDF, causing the parsing process to terminate prematurely. As a result, the parser is unable to complete the full analysis of the document, which further disrupts the extraction of PDF features.

\begin{table}[]
 \centering
 \begin{small}
  \caption{Parser performance comparison results.}
  \vspace{-4mm}
  \label{tab:parse_results}
  \begin{tabular}{llll}
    \toprule
    Parser & Language & Version & Errors\\ 
    \midrule
    pdfrw & Python & 2018 & 4640 \\
    poppler & C++ & 2024 & 632 \\
    pdfminer & Python & 2023 & 937 \\
    MuPDF & C & 2024 & 238 \\
    borb & Python & 2024 & 5971 \\
    QPDF & C++ & 2024 & 592\\
    Poir (Ours) & Python & 2024 & \textbf{0}\\
  \bottomrule
\end{tabular}
 \end{small}
\vspace{-3mm}
\end{table}

Poir has the highest tolerance among these baseline parsers. 
Unlike other Python-based parsers, Poir does not rely solely on the cross-reference table to retrieve objects. When there are errors or incorrect references in the cross-reference table, these Python-based parsers fail to correctly retrieve object references. Poir addresses this issue by scanning each object and performing error correction. Our approach minimizes the failure or incompleteness of feature extraction due to parser errors during the extraction process, demonstrating the highest tolerance for poorly formatted files. 
We have made efforts to avoid parsing errors and pave the way for the conversion and pre-training of PDFObj IR.

\subsection{Evaluation of Embedding Models}
\label{pretrained_eval}

\begin{table}[]
 \centering
  \caption{Prediction accuracy of pre-trained PDFObj2Vec.}
  \begin{small}
  \vspace{-4mm}
  \label{tab:emb-eval}
  \begin{tabular}{lcc}
    \toprule
     Embedding Schemes   & Corpus size = 20k & Corpus size = 65k \\  \midrule
     Word2Vec & 0.7395 & 0.6745 \\
     PV-DM & 0.8857 & 0.8159\\ 
     BERT  & 0.9802 & 0.9796     \\ 
  \bottomrule
\end{tabular}
\vspace{-3mm}
\end{small}
\end{table}

We used two size of corpus to pre-train PDFObj2Vec: one consisting solely of the baseline dataset with 20k samples, and another combining both the baseline and extended datasets into a larger corpus of 65k samples.
The minimum word frequency was set to 1, resulting in vocabulary sizes of 5,554 and 32,347, respectively, to ensure coverage of less frequent malicious terms. Both corpus sizes were split into training, testing, and validation sets in a 7:2:1 ratio.

Due to the differing training objectives of the three embedding models, we devised a unified evaluation strategy. Specifically, we designed cloze tasks on the validation set by randomly generating sentences with missing words and having the models predict these missing words. All models were trained for 100 epochs, and we observed convergence after 20 epochs; therefore, evaluations were based on the performance at the final epoch. The experimental results, presented in Table~\ref{tab:emb-eval}, show that BERT achieved the highest predictive accuracy for both corpus sizes. Interestingly, all three embedding schemes exhibited better performance on the smaller corpus than on the larger one. Word2Vec and PV-DM experienced a drop in accuracy of approximately 6.5\% to 6.8\% when transitioning from the smaller to the larger corpus. This decline can be attributed to the fact that the vocabulary size of the larger corpus is six times that of the smaller corpus, making the prediction task more complex. In contrast, BERT demonstrated superior and consistent performance, effectively overcoming this issue.

\subsection{Evaluation of PDF Malware Classification}
In this subsection, we evaluated the performance of PDFObj2Vec, with different pre-training schemes, corpus sizes, and embedding modes, in the downstream task of PDF malware classification.

\vspace*{2pt}
\noindent \textbf{Classification Experiment Setup }
Following the settings of prior research~\cite{chen2020training, tong2019improving}, we splited the baseline dataset into training and test sets with a 7:3 ratio. 
We embedded the ORG nodes of the samples using various modes and schemes of PDFObj2Vec to generate AORG.
Subsequently, we trained the GIN classifiers on the baseline training set, then we test and evaluate these GIN classifiers on both the baseline test set and the extended dataset. 
The comprehensive experimental results are summarized in Table~\ref{tab:classfication1}.

\vspace*{2pt}
\noindent \textbf{Impact of Pre-Trained Schemes and Corpus Size on Performance }
In the pre-trained mode, regardless of corpus size (20k or 65k), BERT demonstrated the best overall performance, achieving the highest metrics across both the baseline and extended datasets. As the corpus size increased from 20k to 65k, all three schemes showed improvements in their metrics on the baseline dataset.
However, performance on the extended dataset varied. Word2Vec and BERT showed notable gains: Word2Vec\textsc{-65k} improved accuracy by 2.02\% over Word2Vec\textsc{-20k}, and BERT\textsc{-65k} by 2.22\% over BERT\textsc{-20k}. In contrast, PV-DM\textsc{-65k} showed a drop in accuracy compared to PV-DM\textsc{-20k}. 
Experimental results show that a larger pre-training corpus enhances the BERT scheme’s ability to represent PDFObj IR, thereby improving the generalization of the GIN classifier. On the extended dataset, both true positive rate (TPR) and true negative rate (TNR) improved, with TNR much higher than those of the Word2Vec and PV-DM schemes. TPR and TNR present each classifier’s ability to correctly identify both malicious and benign samples. This indicates that the BERT-integrated GIN classifier offers superior classification and generalization performance, particularly for benign samples.

Additionally, we evaluated downstream task performance using intermediate models from different stages of the pre-training process, with results shown in Figure~\ref{fig:all-epochs-ginacc} in Appendix~\ref{diff_epochs_ginacc}. For the BERT scheme, performance gradually stabilized as training progressed and the model converged. In contrast, Word2Vec and PV-DM exhibited less stable, indicating a more erratic convergence.
BERT’s consistent outperformance can be attributed to its bidirectional encoder architecture and pre-training objectives, which enable it to capture complex semantic relationships and model structural dependencies among PDF objects in the ORG structure.

\begin{table}[]
  \caption{PDF malware classification results.}
 \vspace{-4mm}
  \label{tab:classfication1}
  \begin{small}
  \begin{center}

  \begin{tabular}{l|lll|lll}
    \toprule
          Classifiers /   & \multicolumn{3}{c|}{Baseline Dataset (\%)} & \multicolumn{3}{c}{Extended Dataset (\%)}\\ 
     Emb. Schemes                           & Acc & TPR & TNR & Acc & TPR & TNR \\ \midrule
     RTC~\cite{chen2020training}            & 99.15 & 99.97 & 98.11 & 85.15 & 97.04 & 74.72 \\
     CFTC~\cite{tong2019improving}          & 97.58 & 99.75 & 95.04 & 92.04 & 97.31 & 87.41 \\ \midrule
    
    Word2Vec\textsc{-20k}                   & 99.67 & 99.57 & 99.78 & 93.00 & 96.94 & 89.54\\
    PV-DM\textsc{-20k}                      & 99.79 & 99.71 & 99.81 & 93.29 & 97.26 & 89.80 \\ 
    BERT\textsc{-20k}                       & 99.90 & 99.97 & 99.89 & 94.40 & 97.82 & 91.40 \\ \midrule
    Word2Vec\textsc{-65k}                   & 99.82 & 99.68 & 99.85 & 95.02 & 97.65 & 92.72 \\ 
    PV-DM\textsc{-65k}                      & 99.84 & 99.94 & 99.74 & 91.71 & 97.45 & 86.68 \\
    BERT\textsc{-65k}                       & 99.93 & 99.94 & 99.93 & 96.62 & 98.14 & 95.23 \\ \midrule
    
     BERT Base                              & 99.84 & 99.83 & 99.85 & 92.67 & 97.98 & 88.02\\
      CodeT5                                & 99.89 & 99.91 & 99.85 & 94.56 & 98.34 & 91.24 \\ 
       text-emb.-3                          & 99.99 & 99.99 & 99.99 & 97.31 & 98.33 & 96.42 \\  
  \bottomrule
\end{tabular}
\end{center}
\end{small}
\vspace{-5mm}
\end{table}

\vspace*{2pt}
\noindent \textbf{General Embedding Modes }
We evaluated three widely-used, general embedding models: BERT Base, CodeT5, and text-embedding-3, on the PDFObj IR to assess the effectiveness of applying general embeddings directly to classification. The results are presented in the last three rows of Table~\ref{tab:classfication1}.
The experimental results show that among the models tested on PDFObj IR, text-embedding-3 achieves the best performance in the downstream task of PDF malware classification. This can be attributed to its pre-training on a large-scale corpus, which provides it with numerous parameters that enhance its capability. However, text-embedding-3 is computationally expensive and cannot be deployed locally. In contrast, both BERT Base and CodeT5 are more cost-effective and can be deployed locally. Additionally, since CodeT5 is specifically pre-trained for programming languages, and PDFObj IR represents program-like structures, CodeT5 outperforms the general BERT model in PDF malware classification tasks.
In the pre-training mode of PDFObj2Vec, the BERT scheme, which is specifically pre-trained for PDFObj IR, outperforms both the general BERT Base and CodeT5 models in downstream tasks. Although it slightly lags behind text-embedding-3 in overall accuracy, it performs better against adversarial attacks, a point we will discuss in the next subsection.
Furthermore, we also explored the performance of directly embedding the raw content of PDF objects using these general embedding models. We evaluated their classification performance under the same experimental setup, and we will discuss these results in our ablation study (see $\S$\ref{ablation_study}).

\vspace*{2pt}
\noindent \textbf{Compared to Existing Classifiers }
Although RTC and CFTC achieve lower accuracy than our proposed classifier on the baseline dataset, their adversarial robustness is second only to BERT\textsc{-65k}, which we will discuss in the next subsection. 
On the extended dataset, RTC shows a significant drop in accuracy, while CFTC maintains a moderate level of performance, comparable to Word2Vec and PV-DM. Both RTC and CFTC experience only slight decreases in TPR, remaining around 0.97. However, their low accuracy is mainly due to a sharp decline in TNR, indicating a significant increase in the false positive rate (FPR). This due to that these methods overemphasize detecting adversarial malicious samples, at the expense of accurately classifying benign samples.
In contrast, our method demonstrates better generalization in identifying benign samples, which are more diverse in the extended dataset compared to the baseline.

\subsection{Adversarial Attack}
\label{adv_analysis}

\begin{figure*}
    \centering
    \includegraphics[width=1\linewidth]{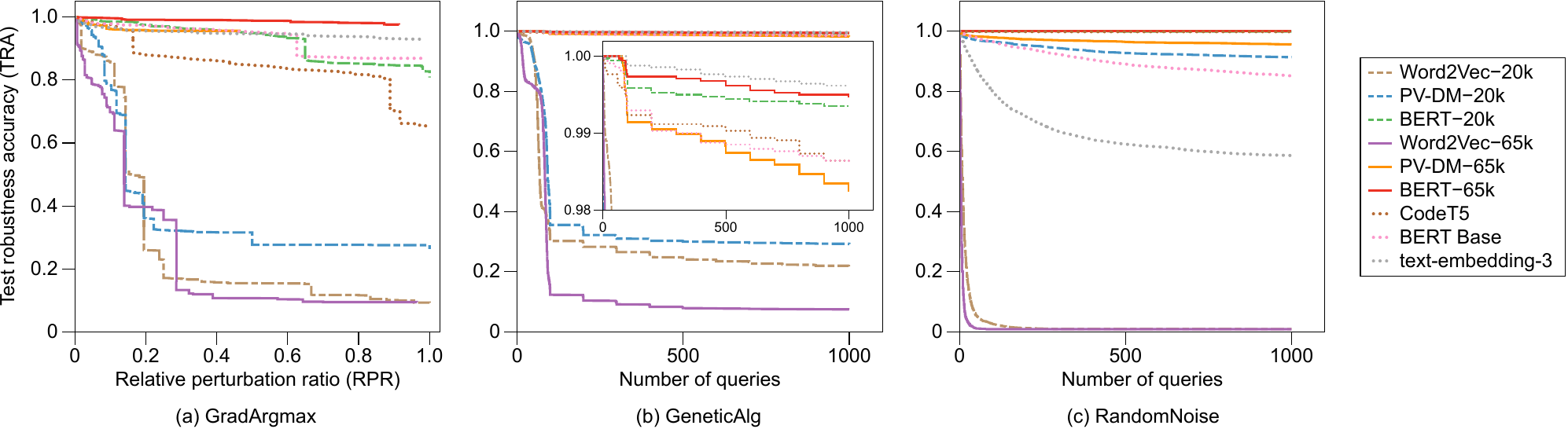}
    \vspace{-6mm}
    \caption{Result of gradient-based attack (a), genetic algorithm-based attack (b), and random noise attack (c).}
    \label{fig:adv_all}
    \vspace{-2mm}
\end{figure*}

\begin{figure}
    \centering
    \includegraphics[width=1\linewidth]{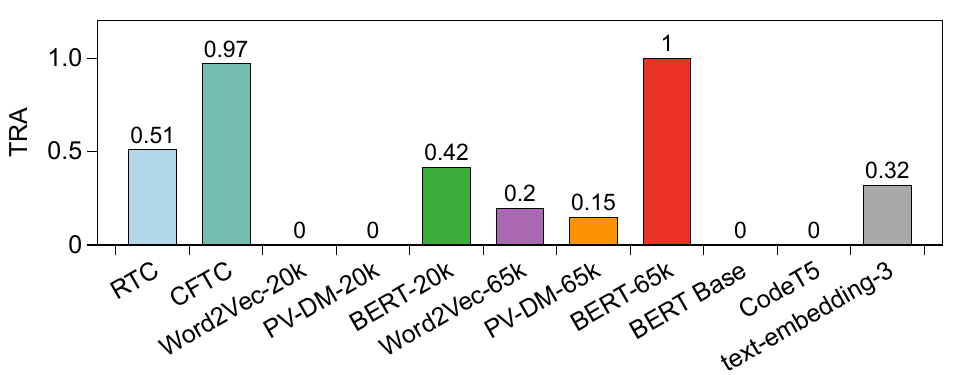}
     \vspace{-7mm}
    \caption{Result of reverse mimicry attack.}
    \label{fig:reverse_mimicry}
    \vspace{-2mm}
\end{figure}

In this subsection, we evaluate the robustness of our proposed PDF malware classifier against four distinct types of adversarial attacks: gradient-based attack, genetic algorithm-based attack, random noise attack, and reverse mimicry attack. We present a summary of these adversarial attacks in Table~\ref{tab:adv_summary} in Appendix~\ref{app:additional_adv_setup}. The primary aim of this evaluation is to understand how different attack strategies affect the model’s performance and to assess the effectiveness of the model’s design in enhancing robustness. Adversarial robustness was assessed using test robustness accuracy (TRA), which measures the proportion of samples that the model can still classify correctly given a specific test input.

\vspace*{2pt}
\noindent \textbf{Gradient-Based Attack }
We adapted gradient-based white-box attack method (GradArgmax) from graph classification tasks~\cite{dai2018adversarial,wan2111adversarial} for PDF classification based on AORG. GradArgmax attack uses classifier gradient information to generate adversarial samples, aiming to reduce the model’s confidence in classifying the target sample. This approach evaluates the robustness of the integrated PDFObj2Vec graph classifier under fine-tuned adversarial examples.
In this scenario, we assume the attacker has full access to the model’s gradient information, with no perturbation distance constraints. The attack budget is set to 1000, consistent with other graph attack study~\cite{wan2111adversarial}. The goal is to identify which graph edges most influence the target classification by evaluating their effect on the loss function using greedy strategy. Specifically, for each candidate edge, its gradient value is computed: if negative, the edge is considered for removal; if positive, it is considered for addition. 
Since the number of nodes and edges in AORG varies, smaller graphs may require only slight perturbations, while larger graphs need more significant ones. To address this, we introduce the relative perturbation ratio (RPR), defined as the ratio of perturbations to the maximum number of edges in the AORG.

We computed the RPR for adversarial samples generated by the GradArgmax attack and plotted the RPR-TRA curve, shown in Figure~\ref{fig:adv_all}(a). A higher RPR at the same TRA level indicates that more perturbations are needed. The results show that BERT\textsc{-65k}, PV-DM\textsc{-65k}, and text-embedding-3 achieved the best TRA performance, with scores of 0.98, 0.95, and 0.93, respectively. Notably, PV-DM\textsc{-65k} had a lower RPR than BERT\textsc{-65k} and text-embedding-3 at similar TRA levels, indicating that it requires more perturbations to be bypassed. Among the models, Word2Vec and PV-DM\textsc{-20k} showed a sharp decline in TRA when RPR < 0.2, showing the weakest performance.

\vspace*{2pt}
\noindent \textbf{Genetic Algorithm-Based Attack }
We implemented a genetic algorithm-based black-box attack (GeneticAlg)~\cite{dai2018adversarial}, where the attacker has no access to the model’s internal details, only the output confidence scores. GeneticAlg leverages a genetic algorithm to optimize the structure of the input graph, effectively bringing the sample closer to the model’s decision boundary. This approach is used to evaluate the defensive capabilities of our proposed graph classifier against adversarial samples generated through complex search strategies.
And, GeneticAlg consists of five key components: population, fitness function, selection, crossover, and mutation. The attack involves edge flipping, node injection, and deletion. Each generation evolves based on the fitness of the mutated samples, guiding the direction for subsequent generations. Following the experimental setup in \cite{dai2018adversarial}, we set the attacker query limit to 1000, the population size to 100, and the number of generations to 10. The results, shown in Figure\ref{fig:adv_all}(b), indicate that Word2Vec\textsc{-20k} and PV-DM\textsc{-20k} performed the worst in terms of TRA, while the other models maintained TRA values above 0.98, with text-embedding-3 and BERT\textsc{-65k} performing the best. Overall, the results suggest that our proposed classifier demonstrates strong defenses against the GeneticAlg attack.

\vspace*{2pt}
\noindent \textbf{Random Noise Attack }
We extended node classification attack methods from \cite{ma2020towards} to perturb node features for whole-graph classification. Specifically, we injected Gaussian noise into the node features of the AORG graph to evaluate the model’s resistance to random perturbations, thereby testing its robustness under noisy conditions.
To select the attack nodes, we employed degree centrality, as it provides a reasonable baseline for graph-based attacks~\cite{newman2018networks}. High-degree nodes were prioritized because they play a crucial role in information propagation and are more likely to significantly impact the overall graph structure. Gaussian noise was then introduced to perturb the features of these nodes. In this black-box attack scenario, the attacker only has access to the model’s binary output (0 or 1) and is limited to a maximum of 1000 queries.
The experimental results, shown in Figure~\ref{fig:adv_all}(c), indicate that Word2Vec exhibited the lowest robustness under these conditions. In contrast, BERT\textsc{-65k} and BERT\textsc{-20k} showed the highest performance, with TRA values of 1 and 0.999, respectively. It suggests that BERT scheme are more resilient to adversarial noise, particularly with respect to node feature perturbations. The TRA of text-embedding-3 showed a significant decline under the random noise attack, compared to the GradArgmax and GeneticAlg attacks. This indicates that its embeddings are highly susceptible to noise interference.

\vspace*{2pt}
\noindent \textbf{Reverse Mimicry Attack }
We used the reverse mimicry adversarial attack~\cite{maiorca2013looking} for evaluation, which has been evaluated in previous state-of-the-art robust PDF malware classifiers~\cite{chen2020training,tong2019improving}. This black-box attack is independent of specific classifiers or features, making it suitable for evaluating our proposed graph classifier as well as the RTC and CFTC classifiers, which employ different models and features. Unlike other attacks, reverse mimicry directly manipulates the sample space by injecting malicious payloads into benign samples, thereby generating realizable adversarial examples designed to evade classification boundaries. This approach enables a rigorous evaluation of the model's robustness against intentionally disguised adversarial samples.

We followed the adversarial evaluation settings from \cite{chen2020training,tong2019improving}, using 500 seed samples to create adversarial samples. A Cuckoo sandbox~\cite{Cuckoo} was set up to test the adversarial examples. If they generated the expected network communication metrics, they were deemed legitimate. The experimental results, shown in Figure~\ref{fig:reverse_mimicry}, reveal that with a 20k corpus size, only BERT\textsc{-20k} demonstrated some resistance to adversarial samples. 
However, with a 65k corpus size, all three PDFObj2Vec pre-training models successfully detected adversarial samples. BERT\textsc{-65k} achieved a TRA of 1, while the other models had TRA values below 0.2. 
These findings suggest that BERT\textsc{-65k} is particularly effective at capturing the semantic information of nodes. When malicious payloads are injected into benign samples, the corresponding malicious nodes are also present in the adversarial samples. BERT\textsc{-65k} accurately represents these embeddings and, through the node propagation and aggregation process in GIN, successfully identifies the adversarial samples.
Notably, RTC and CFTC achieved a TRA of over 0.5 but at the cost of significantly reduced classification accuracy, as shown in Table~\ref{tab:classfication1}. Specifically, CFTC achieved only 97.58\% accuracy on the baseline dataset. And both CFTC and RTC exhibited significantly lower TNR compared to our proposed classifier. This indicates a higher FPR for CFTC and RTC, for instance, CFTC reached an FPR of 4.96\% on the baseline dataset and an FPR of 12.59\% on the extended dataset, substantially higher than that of BERT\textsc{-65k}.

\begin{table}[]
 \centering
  \caption{TRA under GradArgmax and GeneticAlg attacks at maximum query budget.}
  \begin{small}
  \vspace{-2mm}
  \label{tab:tra_compared}
  \begin{tabular}{lcccc}
    \toprule
     Attack     & \multicolumn{4}{c}{Classifiers / Emb. Schemes}   \\
     Method     & RTC    & CFTC  & BERT\sc{-65k} & text-embedding-3\\ \midrule
     GradArgmax & 0 & 0.94 & 0.98 & 0.93\\
     GeneticAlg & 0.81 & 0.71& 0.99 & 0.99\\
  \bottomrule
\end{tabular}
\vspace{-5mm}
\end{small}
\end{table}

\vspace*{2pt}
\noindent \textbf{Results Analysis }
We evaluated our proposed classifier using four adversarial attack methods that target different space. 
Among these, the reverse mimicry attack operates in the sample space and simulates a realistic black-box scenario. In contrast, GradArgmax, GeneticAlg, and RandomNoise are feature-space attacks that, while less realistic, are widely used to stress-test classifiers under strong assumptions~\cite{dai2018adversarial,wan2111adversarial,ma2020towards,chen2020training}. These attacks help reveal how vulnerable a model may be when its decision boundaries are exploited, offering valuable insights into worst-case robustness.

The graph features used in our proposed GIN classifier differ fundamentally from the Hidost features employed by RTC and CFTC. Consequently, the aforementioned feature-space attacks (GradArgmax, GeneticAlg, and Random Noise) are not directly applicable to RTC and CFTC, as they are designed to perturb graph nodes and structures. To enable a fair comparison, we implemented GradArgmax and GeneticAlg attacks specifically tailored for Hidost features using bit-flipping as the basic operation with the same attack intensity. The TRA results are shown in Table~\ref{tab:tra_compared}, with TRA-versus-perturbation curves shown in Figure~\ref{fig:adv_rtcc_ftc} in Appendix~\ref{app:additional_adv_setup}. Since Hidost features are binary (0 or 1), adding random noise would produce non-binary values, invalidating the feature representation; thus, RandomNoise attacks were not evaluated for these classifiers. The results indicate that under GradArgmax, RTC’s TRA drops to 0 while CFTC achieves a TRA of 0.94. Under GeneticAlg, both RTC and CFTC maintain relatively high TRA, though their robustness remains lower than that of the classifiers based on BERT\textsc{-65k} and text-embedding-3.

Overall, classifiers utilizing pre-trained embedding schemes exhibit stronger adversarial robustness than those using general embeddings. In particular, the graph classifier integrated with BERT\textsc{-65k} demonstrates the highest adversarial resilience. This can be attributed to the architecture of BERT and the design of the NOP and MLM training tasks, which not only enable BERT to learn node semantic embeddings but also capture the contextual relationships between nodes. As a result, the BERT\textsc{-65k} integrated graph classifier achieves superior robustness to both graph structure perturbations and node feature disturbances.

\begin{table}[]
 \centering
  \caption{Results of ablation study. $R$ refers to embeddings applied directly on raw content. }
  \begin{small}
  \vspace{-2mm}
  \label{tab:ablation_result1}
  \begin{tabular}{lcccr}
    \toprule
    Emb. & \multirow{2}{*}{Classifiers} & Baseline & Extended & Adv Samples \\
    Schemes &  & Acc(\%)  & Acc(\%)  & \multicolumn{1}{c}{TRA}\\ \midrule
    BERT$^R$\textsc{-20k}  & GIN    & 99.48 & 92.33    & 0.42 $\rightarrow$ 0 \\ 
    BERT Base$^R$ & GIN & 98.55 & 85.11     & 0 $\rightarrow$ 0  \\
    CodeT5$^R$  & GIN  & 99.71 & 91.75    & 0 $\rightarrow$ 0  \\ \midrule
    BERT\textsc{-20k} & DNN & 99.80 & 89.85 & 0.42 $\rightarrow$ 0 \\
    PV-DM\textsc{-65k} & DNN & 99.66 & 92.84 & 0.15 $\rightarrow$ 0 \\
    BERT\textsc{-65k} & DNN & 99.92 & 96.10 & 1 $\rightarrow$ 0 \\
    text-emb.-3 & DNN & 99.82 & 89.75 & 0.32 $\rightarrow$ 0 \\
  \bottomrule
\end{tabular}
\vspace{-5mm}
\end{small}
\end{table}

\subsection{Ablation Study}
\label{ablation_study}

In this section, we conduct ablation studies to evaluate the impact of PDFObj IR and ORG on the performance of PDF malware classification. Specifically, we design two experiments: 1) removing PDFObj IR and using standard BERT pre-training and general embedding schemes on the raw content of PDF objects; and 2) removing the ORG structure and GIN classifier, replacing them with a conventional deep neural network (DNN) classifier.

\vspace*{2pt}
\noindent \textbf{Impact of Embedding Raw Content on Classification }
In this experiment, we first pre-trained the raw content of the PDF object, shown in Figure~\ref{fig:pdf_background}(b), using the standard BERT pre-training method with default preprocessing and tokenization configurations. This pre-training was performed on the baseline dataset (with a corpus size of 20k samples). Then, we applied the general embedding schemes from $\S$\ref{general_emb} to embed the raw content of PDF objects, obtaining node embeddings. We then evaluated their performance on the PDF malware classification task, maintaining the same GIN model and training settings as described in $\S$\ref{gnn_classify}. The results, presented in the first three rows of Table~\ref{tab:ablation_result1}, indicate that the BERT$^R$ scheme, pre-trained on raw content, performs worse than the BERT scheme pre-trained on PDFObj IR (as shown in Table~\ref{tab:classfication1}). Specifically, with a corpus size of 20k, the accuracy on the baseline dataset decreased by 0.42\%, and on the extended dataset, it dropped by 2.07\%. Furthermore, the results show that directly applying general embedding schemes to raw content performs worse than applying them at the PDFObj IR level. For instance, the BERT Base$^R$ scheme demonstrated a significant performance drop: compared to BERT Base in Table~\ref{tab:classfication1}, accuracy on the baseline and extended datasets decreased by 1.29\% and 8.56\%, respectively. Similarly, CodeT5$^R$ saw accuracy reductions of 0.18\% and 2.84\%, respectively. Detailed results are provided in Appendix~\ref{app:removal}.

Additionally, we tested the resistance of these schemes to adversarial samples generated by reverse mimicry. None of these schemes showed resistance to adversarial samples. 
The TRA of the BERT\textsc{-20k} scheme on PDFObj IR against the reverse mimicry attack dropped from 0.42 to 0. The results from our experiments highlight the pivotal role of PDFObj IR in enhancing the performance of PDF malware classification tasks. Compared to raw content, PDFObj IR provides a more abstract and refined representation of the PDF object content by removing irrelevant strings such as symbols, special characters, and escape sequences. In contrast, embeddings applied directly to raw content are influenced by these extraneous strings, which are incorporated into the node embeddings, leading to inaccuracies in classification. The pre-training process on PDFObj IR enables embedding schemes to more precisely capture the semantics of PDF objects, ultimately improving classification performance.

\noindent \textbf{Impact of ORG Structure on Classification }
To evaluate the impact of graph structures on classifier robustness, we removed the ORG structure and GIN classifier. Given the variability in the number of objects across PDF samples, we represented each PDF by averaging the semantic vectors of all objects and used this representation as input to a DNN classifier. We selected four PDFObj2Vec schemes that demonstrated some adversarial robustness, as discussed in $\S$\ref{adv_analysis}. 
We trained the DNN classifier on the baseline dataset and assessed its performance. 
The experimental results highlight the critical importance of the ORG structure in enhancing both the performance and robustness of PDF malware classification. When the ORG structure was removed and a DNN classifier was used with semantic vectors of PDF objects, the classification accuracy on both the baseline and extended datasets declined. Furthermore, the DNN classifiers completely failed to exhibit resistance to adversarial samples, rendering them unable to detect such attacks effectively.

The strength of ORG lies in its ability to model the inter-object relationships and structural dependencies within a PDF. Unlike simple vector averaging, which treats objects as independent and unstructured entities, ORG represents PDFs as a graph, where nodes capture the semantic properties of objects and edges encode their reference relationships. 
This structured representation enables the classifier to capture global patterns and contextual dependencies that are crucial for distinguishing between benign and malicious PDFs, especially in complex attack scenarios. Additionally, ORG 
is well-suited for integration with graph neural networks, such as the GIN classifier, which are inherently designed to learn from graph-structured data.
By combining semantic and structural modeling, this approach allows the classifier to achieve high accuracy while robustly defending against adversarial attacks, as evidenced by the notable performance gains in the experiments.

\section{Discussion \& Conclusion}

\noindent \textbf{Obfuscation }
Obfuscation is a common tactic in malware to evade detection, and similar techniques, such as keyword obfuscation, are used in PDF malware to bypass detection mechanisms~\cite{obs_pdfs_sentinelone,lu2013obfuscation,Pires_medium}. For example, replacing Name \texttt{/URI} with \texttt{/\#55RI}, prevents classifiers that rely on keyword statistical features from obtaining correct statistical features and also hinders analyzers that use Hidost path features from extracting the correct path features. Our method mitigates such obfuscation by retaining the obfuscated keyword data in the PDFObj IR corpus, allowing us to learn the obfuscated semantics. Additionally, we integrate the structural features of ORG to classify PDF malware, enhancing the precise detection of obfuscated PDF malware.

\vspace*{2pt}
\noindent \textbf{Concept Drift }
Over time, the effectiveness of traditional machine learning models trained on outdated datasets tends to decline~\cite{jordaney2017transcend,yang2021cade,barbero2022transcending}. 
As multimedia evolves, benign PDFs are advancing faster than malicious ones, altering their characteristics significantly. Consequently, SOTA classifiers, including RTC~\cite{chen2020training} and CFTC~\cite{tong2019improving}, struggle to accurately identify new benign samples of the extended dataset, leading to lower accuracy, as shown in Table~\ref{tab:classfication1}. 
Since PDFObj2Vec is trained on PDFObj IR, a fundamental representation of PDF objects, the basic changes in PDFObj IR are minimal, regardless of variations in the PDF structure. 
Therefore, our proposed method is minimally affected by concept drift.
The experimental results presented in Table~\ref{tab:classfication1}  demonstrate that our proposed approaches, which leverage PDFObj IR, not only outperform methods that do not utilize this representation but also surpass SOTA classifiers across all metrics on the most recent extended dataset.

\vspace*{2pt}
\noindent \textbf{Conclusion }
In this paper, we present PDFObj IR, an intermediate representation framework specifically tailored for PDF analysis. Leveraging PDFObj IR, we constructed an Object Reference Graph (ORG) and developed a method for node semantic extraction, termed PDFObj2Vec. Additionally, we designed a classifier based on the Graph Isomorphism Network (GIN) and evaluated the performance of PDFObj2Vec-GIN across multiple datasets and against adversarial samples. The results demonstrate that PDFObj2Vec-GIN achieves exceptional classification performance and exhibits significant adversarial robustness, highlighting the effectiveness of our proposed analytical framework.

\section*{Acknowledgment}

We thank all anonymous reviewers for their valuable comments to improve this paper. 
This work is supported by the National Nature Science Foundation of China under Grant No.62172308, 62272351, 61972297, and 62172144. Jiang Ming is supported by NSF grants 2312185 \& 2417055, Google Research Scholar Award, and Tulane COR Fellowships.







{
\bibliographystyle{ACM-Reference-Format}
\bibliography{ref}
}

\newpage
\appendix
\section*{Appendix}
\renewcommand{\thesubsection}{\Alph{subsection}}

\subsection{Basic Type Description}
\label{basic_desc}
PDF objects' base value types are described in Table~\ref{tab:valuetype}.

\begin{table}[h]
  \centering
  \caption{Basic type description.}
  \label{tab:valuetype}
  \begin{tabular}{lp{6cm}}
    \toprule
    Type & Description \\ 
    \midrule
    Numeric & Numeric object, e.g., 2 and 3.14;\\
    String & It enclosed in parentheses or angle brackets. The former is the literal string, and the latter is the hex string, e.g., \texttt{(en-US)}, \texttt{<3DA7>} \\
    Name & It used as key in dictionaries and starting with \texttt{/}, e.g., \texttt{/Pages} and \texttt{/Type}\\
    Boolean & It represented by the keywords \texttt{True} and \texttt{False} \\
    Null & It indicated by the keyword \texttt{null} \\
  \bottomrule
\end{tabular}
\end{table}

\subsection{Error Handling and Correction}
\label{other_error_process}
Apart from the three main errors discussed in $\S$\ref{error_process}, there are other errors that we have identified. These additional errors and their handling measures are listed in Table~\ref{tab:appd_error_handling}.

\begin{table*}[]
  \centering
  \caption{PDF parsing errors and handling measures.}
  \label{tab:appd_error_handling}
  \begin{tabular}{lp{6cm}p{9cm}}
    \toprule
    No. & Error Description & Handling Measures \\ 
    \midrule
    E4 & Missing version number & Ignore version number, do not match PDF Header \\
    E5 & Missing "/Root" object pointer in the Trailer dictionary & Ignore Root object, linearly scan the Body \\ 
    E6 & Missing "\%\%EOF" keyword at the end & Ignore the "\%\%EOF" keyword, use the "trailer" keyword to locate the start of the Trailer, and the next keyword to locate the end of the Trailer\\
    E7 & Missing "/Length" attribute in the stream object dictionary &
Ignore the "/Length" attribute and calculate the length of the stream data based on the "endstream" keyword or the next keyword. \\
E8 & Xref table format error or invalid object offsets &
Linearly scan the objects in the body and manually repair the reference relationships\\
  \bottomrule
\end{tabular}
\end{table*}

\subsection{Details of Pre-trained PDFObj2Vec }
\label{word2vec_pvdm_details}

\noindent \textbf{BERT Scheme} 
We designed two training tasks for BERT scheme: MLM and NOP.
The loss $\mathcal{L}_{MLM}$ for the MLM task is the cross-entropy loss:
$$\mathcal{L}_{MLM} = - \sum_{IR_i \in m(Seq_{IR})} log P(\hat{IR_i} \mid Seq_{IR}) $$
where $m(Seq_{IR})$ denotes the set of tokens that are masked.
And the loss $\mathcal{L}_{NOP}$ for the NOP task is also the cross-entropy loss:
$$\mathcal{L}_{NOP} = - log P (y \mid obj_1, obj_2) $$
The loss function of the BERT scheme is the combination of $\mathcal{L}_{NOP}$ and $ \mathcal{L}_{MLM}$.
$$\mathcal{L} = \mathcal{L}_{MLM} + \mathcal{L}_{NOP}$$

\begin{figure}[]
    \centering
    \includegraphics[width=0.88\linewidth]{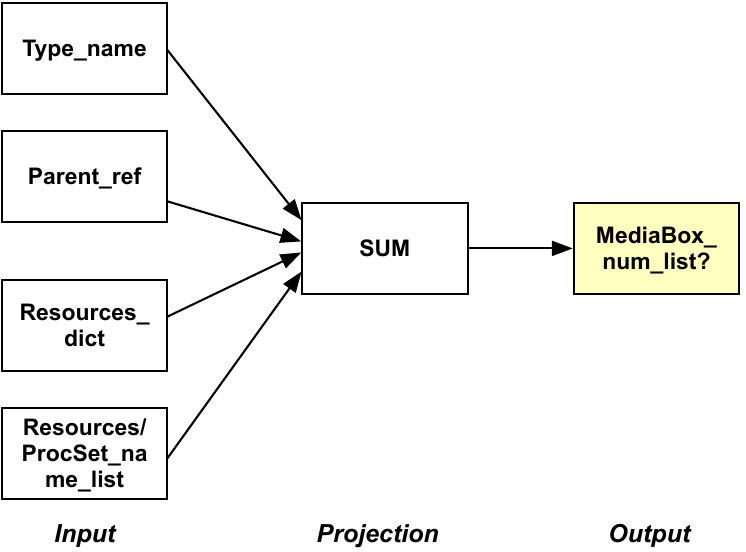}
    \vspace{-2mm}
    \caption{The training task for Word2Vec.}
    \label{fig:word2vec_cbow}
    \vspace{-2mm}
\end{figure}

\noindent \textbf{Word2Vec-Based Scheme} 
For a sentence structure like \hytt{Type\_name} \hytt{Parent\_ref} \hytt{MediaBox\_num\_list} \hytt{Resources\_dict} \hytt{Resources/P-rocSet\_name\_list} \hytt{Resources/XObject\_DICT...} , suppose we select \hytt{MediaBox\_num\_list} as the target word, and the context window size is 2. This means we will use the two words before and after this word as the context, as shown in Figure~\ref{fig:word2vec_cbow}. Thus, the context words are: [\hytt{Type\_name} \hytt{Parent\_ref} \hytt{Resources\_dict} \hytt{Resources/ProcSet\_name\_list}]. The overall semantics of the context are represented by calculating the average of the context word vectors. This context vector is then fed into a softmax classifier, which predicts the probability of each word being the target word. 
In $\S$\ref{evaluation}, our implementation of Word2Vec consists of two embedding layers, one for learning the embeddings of the center words and the other for the context words, along with a Dropout layer. We update the word vectors using the cross-entropy loss between the target words and the model's predictions. The optimizer used is Stochastic Gradient Descent (SGD), with a learning rate of 5e-4.

\begin{table}[]
 \centering
  \caption{Hyperparameters of BERT scheme of PDFObj2Vec.}
  \label{tab:parameters_obj2vec_bert}
  \begin{tabular}{ll}
    \toprule
    Hyperparameters & BERT \\
    \midrule
    attention\_probs\_dropout\_prob & 0.1\\
    hidden\_activation & gelu\\
    hidden\_dropout\_prob & 0.1\\
    initializer\_range & 0.02\\
    intermediate\_size & 307\\
    layer\_norm\_eps & 1e-12\\
    max\_position\_embedding & 512\\
    num\_attention\_heads & 8\\
    num\_hidden\_layers & 8\\
    pad\_token\_id & 0\\
    position\_embedding\_type  & absolute\\
    transformers\_version  & 4.38.2\\
    type\_vocab\_size  & 2\\
    min\_freq & 1 \\
    hidden\_size & 512 / 1024 \\
  \bottomrule
\end{tabular}
\end{table}

\vspace*{2pt}
\noindent \textbf{PV-DM-Based Scheme} 
PV-DM is based on the Word2Vec concept, where it learns word embeddings by predicting a target word from context words or predicting context words from a target word. However, PV-DM further introduces paragraph-level vectors, allowing the model to capture a broader context. In the PV-DM model, we consider the relationships between words in the same paragraph to be closer than those between words in different paragraphs. The goal is to predict the current word using context words and a paragraph vector. In the example in Figure~\ref{fig:bertinput}, Object A and Object B can be seen as two paragraphs. When the target word is \texttt{Kids\_ref\_list} in object A, the input for PV-DM is
$$Input = [V_{ObjectA}; V_{Type\_name}; V_{Count\_num}]$$
where $V_{ObjectA}$ is the vector for Object A, $V_{Type\_name}$ is the vector for \texttt{Type\_name}, and $V_{Count\_num}$ is the vector for \texttt{Count\_num}. The input is the concatenation of these three vectors, and the objective is to maximize the conditional probability of the target word, given these input vectors. PV-DM learns word and document vector representations by combining the document vector and context word vectors at each prediction step. The optimizer used is SGD with a learning rate of 5e-4.

\subsection{Settings of Pre-trained PDFObj2Vec}
\label{Settings_PDFObj2Vec}
Under the pre-training mode, we implemented three schemes with Pytorch: Word2Vec, PV-DM, and BERT. Some of their hyperparameters are listed in Table~\ref{tab:parameters_obj2vec_bert} and Table~\ref{tab:parameters_baseemd}. Each scheme was trained for 100 epochs, and the batch size is 64.

\begin{table}[]
 \centering
  \caption{Hyperparameters of Baseline Embedding Models.}
  \label{tab:parameters_baseemd}
  \begin{tabular}{ll}
    \toprule
    Hyperparameters & Word2Vec / PV-DM \\
    \midrule
    learning rate & 5e-4\\
    min\_freq & 1 \\
    window\_size & 1 \\
    neg\_count & 5  \\
    embedding\_dimension & 512 / 1024 \\
  \bottomrule
\end{tabular}
\end{table}

\subsection{Details of GIN Classifier}
\label{GIN_Classifier}

We trained our implemented GIN classifier on the baseline dataset using the Deep Graph Library~\cite{gdlai}. The classifier consists of two GIN layers, each with a hidden layer size of 256, employing mean aggregation and the ReLU activation function. 
The update rule for a GIN graph convolution layer can be expressed as:
$$ h_v^{(k+1)} = \text{MLP}^{(k)} \left( (1 + \epsilon^{(k)}) \cdot h_v^{(k)} + \sum_{u \in \mathcal{N}(v)} h_u^{(k)} \right) $$
In this formula, $h_v^{(k)}$ represents the feature vector of node $v$ at the $k$-th iteration or layer,  $\mathcal{N}(v)$  denotes the set of neighbors of node $v$, and  $\epsilon^{(k)}$ is a learnable parameter that can adjust the weighting of the node's features relative to its neighbors. The MLP is applied at each layer $k$, and it is responsible for transforming the aggregated feature vector into the next layer’s node features. 
The input dimension for the linear layer is 256, and the output dimension is 2, corresponding to the probabilities of being classified as malicious or benign. The optimizer used is Adam, with a learning rate of 0.01. The loss function is binary cross-entropy loss, the batch size is 64, and the model was trained for 50 epochs.

\subsection{Settings of Baseline Classifiers}
\label{baselines_classifiers_details}

We reproduced the results of RTC and CFTC using their publicly available models. 
To avoid potential discrepancies caused by retraining the models from scratch, we directly utilized the well-trained model files released by the original authors. Accordingly, we only made necessary adjustments related to environment setup, evaluation code adaptation, and workflow integration to ensure that our reproduced results closely match those reported in the original papers. For instance, RTC achieved an accuracy of 0.9915 on the contagio dataset, which is consistent with the original paper. As for CFTC, the original work did not report the exact accuracy on the contagio dataset, but it did provide an AUC of 0.9982. Our reproduced AUC is 0.9984, which is very close to the original and even slightly better.

\vspace*{2pt}
\noindent \textbf{Robustly Trained Classifier (RTC) } 
RTC is based on a baseline deep neural network (DNN) containing two hidden fully connected layers, each with 200 neurons activated by ReLU and a final layer with Softmax activation. Then, Chen et al. used a symbolic interval analysis method to robustly retrain a neural network classifier with the same model architecture and the same set of hyperparameters as DNN. We adopted the most robust architecture, Robust A+B+E, and it employs the Adam optimizer with a learning rate of 0.01, a batch size of 64, and 50 training epochs.

\vspace*{2pt}
\noindent \textbf{Conserved Features Trained Classifier (CFTC) } 
CFTC is based on a baseline support vector machine classifier (SVC) and identifies seven conserved Hidost path features. These features are then conserved for iterative adversarial training of the SVC using an RBF kernel. The penalty coefficient of the objective function is set at 12, and the coefficient for the kernel function is 0.0025. All other parameters follow the default settings of the SVC class in scikit-learn package.

\subsection{Different Stages of the Pre-Training}
\label{diff_epochs_ginacc}
\begin{figure}[]
    \centering
    \includegraphics[width=1\linewidth]{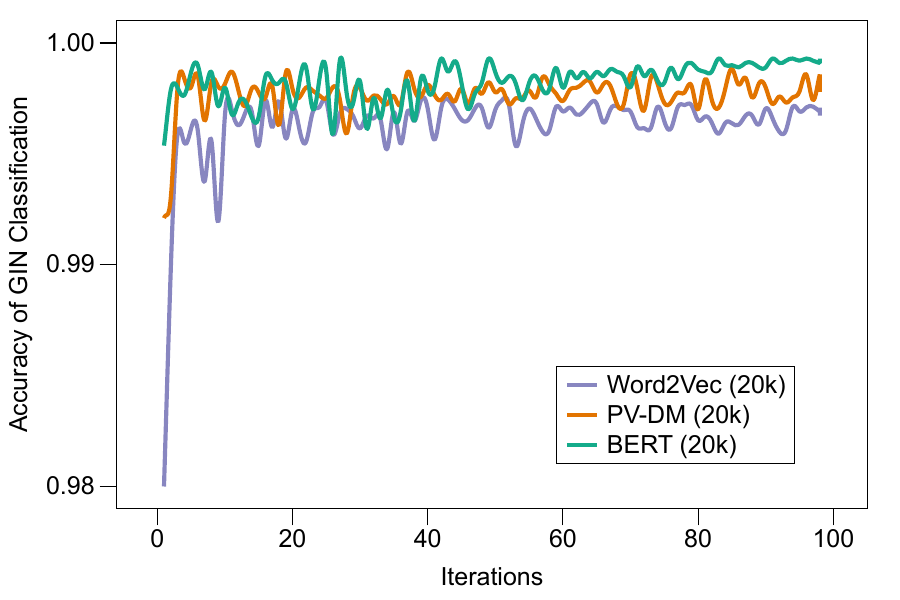}
    \vspace{-5mm}
    \caption{GIN classifier's acccuracy with different PDFObj2Vec's pre-training epochs.}
    \label{fig:all-epochs-ginacc}
    \vspace{-2mm}
\end{figure}

To explore the impact of intermediate models generated at different stages of PDFObj2Vec pre-training on the PDF malware classification task, we applied these intermediate models (corpus size of 20k) to the classification task. We followed the same classification task setup as before, training and evaluating the classifier on the baseline dataset. The experimental results are shown in Figure~\ref{fig:all-epochs-ginacc}. Since all three pre-trained schemes began to converge within the first 20 iterations during training, they quickly reached a high accuracy range in the classification task as well. 
The performance of the BERT scheme in downstream tasks gradually stabilized over the iterations, with reduced fluctuations. In contrast, PV-DM and Word2Vec exhibited larger fluctuations than BERT, showing less stability. This indicates that improvements in pre-training quality can effectively enhance BERT's performance in downstream tasks. However, for Word2Vec and PV-DM, improvements in pre-training quality result in only limited gains in downstream task performance.

\begin{table*}[]
 \centering
  \caption{Summary of adversarial attack settings. }
  \vspace{-3mm}
  \label{tab:adv_summary}
  \begin{center}
  \begin{tabular}{cccccc}
    \toprule
     Attack Method    & Type & Target Space & Access & Interact? & Realisitc? \\  \midrule
     Gradient-Based & White Box    & Feature    & Gradient & \tiny{\Checkmark} & \tiny{\XSolidBrush} \\
     Genetic Algorithm-Based & Black Box    & Feature    & Score  & \tiny{\Checkmark} & \tiny{\XSolidBrush}\\
     Random Noise & Black Box   & Feature   & Label  & \tiny{\Checkmark} & \tiny{\XSolidBrush}\\
     Reverse Mimicry & Black Box    & Sample  & None  & \tiny{\XSolidBrush} & \tiny{\Checkmark}\\
     
     \bottomrule
\end{tabular}
\end{center}
\end{table*}

\subsection{Additional Details of Adversarial Attack}
\label{app:additional_adv_setup}

We summarize the adversarial attacks we use in $\S$\ref{adv_analysis}, as shown in Table~\ref{tab:adv_summary}.
Specifically, the gradient-based attack, genetic algorithm-based attack, and random noise attack are conducted in the feature space, where attackers are assumed to have varying degrees of access to the model's internals (e.g., gradients or confidence scores). These settings are idealized and do not correspond directly to practical attack conditions.
However, these attacks are included for two important reasons. First, they allow us to assess the internal robustness of our model under strong adversarial conditions, thereby providing insight into its behavior in worst-case scenarios. For example, the gradient-based attack assumes white-box access and is designed to expose vulnerabilities in the learned feature space by directly manipulating model-sensitive directions. The genetic algorithm-based attack simulates a black-box adversary with query access to confidence scores, reflecting a more restricted but still potent threat model. Random noise attacks, while non-adaptive, serve as a baseline to evaluate the model's resilience to generic feature perturbations.
Second, such feature-space attacks are commonly adopted in the adversarial robustness studies for stress-testing models under controlled conditions. They are not intended to simulate realistic attacks, but rather to reveal how easily a classifier may be fooled when the attacker can exploit flaws in its learned decision boundaries.

In contrast, the reverse mimicry attack operates in the sample space and simulates a more realistic black-box setting, where an adversary manipulates input-level content to mimic benign samples. This attack better reflects real-world scenarios and complements the others by demonstrating robustness under practical constraints.

\begin{figure}[]
    \centering
    \subfigure[GradArgmax]{
        \includegraphics[width=0.225\textwidth]{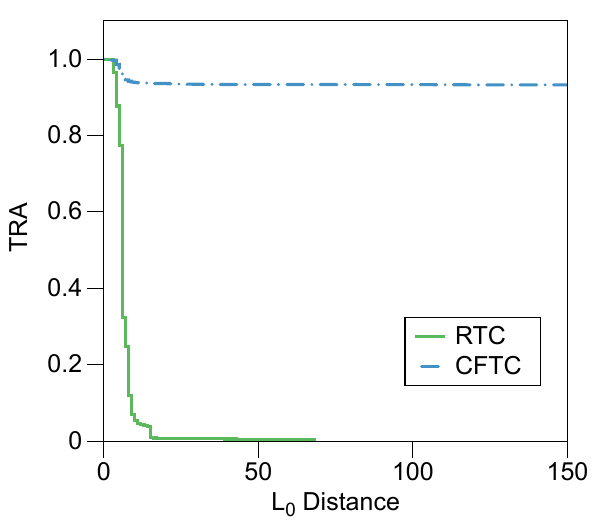}
        \label{fig:adv_grad}
    }
    \subfigure[GeneticAlg]{
        \includegraphics[width=0.225\textwidth]{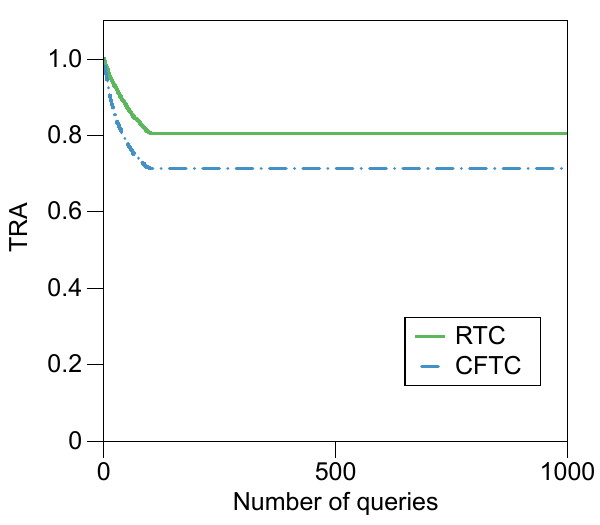}
        \label{fig:adv_gen}
    }
    \vspace{-4mm}
    \caption {Results of RTC and CFTC under GradArgmax and GeneticAlg attacks.}
    \label{fig:adv_rtcc_ftc}
    \vspace{-2mm}
\end{figure}

In addition, we include the TRA–$L_0$ distance curves for RTC and CFTC under GradArgmax and GeneticAlg attacks, as shown in Figure~\ref{fig:adv_rtcc_ftc}. Here, the $L_0$ distance represents the number of perturbed Hidost features required for a successful attack, serving as a measure of perturbation strength. 
This is conceptually similar to the Relative Perturbation Ratio (RPR) discussed in $\S$\ref{adv_analysis}.

\subsection{Detailed Results of Ablation Study}
\label{app:removal}
The detailed results of the ablation study are presented in Table~\ref{tab:classfication_removal_dnn}. Removing PDFObj IR and applying embeddings directly on raw content resulted in a decline in the classifier’s performance. Additionally, we selected models in $\S$\ref{adv_analysis} that exhibited a certain level of adversarial robustness: BERT\textsc{-20k} PV-DM\textsc{-65k}, BERT\textsc{-65k}, and text-embedding-3. Subsequently, we removed the ORG structure and used the average of node vectors for classification, and the classification results are shown in Table~\ref{tab:classfication_removal_dnn}. The results indicate that after removing the ORG structure, the models that originally had adversarial capabilities no longer detected any adversarial samples, suggesting that the ORG structure contributes to improving the classifier's adversarial robustness. Additionally, we observed that following the removal of the ORG structure, apart from BERT\textsc{-65k}, the combinations of other embedding schemes with DNN classifiers experienced a certain degree of decline in classification metrics compared to the combinations with GIN classifiers. This underscores that our trained BERT\textsc{-65k} scheme is more accurate in representing the semantics of PDF objects.

\begin{table*}[]
  \caption{Comparative evaluation results of removal of PDFObj IR and ORG structure.}
 \vspace{-3mm}
  \label{tab:classfication_removal_dnn}
  \begin{center}
  \begin{tabular}{cc|ccc|ccc|c}
    \toprule
    Embedding          & \multirow{2}{*}{Classifiers} & \multicolumn{3}{c|}{Baseline Dataset (\%)} & \multicolumn{3}{c|}{Extended Dataset (\%)} & Adversarial Samples\\
    Schemes         &    & Acc & TPR & TNR & Acc & TPR & TNR  & TRA             \\ \midrule
    BERT$^R$\textsc{-20k}   & GIN & 99.48 & 99.39 & 99.59 & 92.33 & 96.19 & 88.93 & 0 \\ 
    BERT Base$^R$           & GIN & 98.55 & 99.18 & 97.74 & 85.11 & 95.96 & 75.58 & 0 \\
    CodeT5$^R$              & GIN & 99.71 & 99.88 & 99.48 & 91.75 & 96.73 & 87.38 & 0 \\ \midrule
    BERT\textsc{-20k}       & DNN & 99.80 & 99.97 & 99.59 & 89.85 & 98.48 & 82.28 & 0 \\
    PV-DM\textsc{-65k}      & DNN & 99.66 & 99.71 & 99.56 & 92.84 & 97.53 & 88.72 & 0 \\
    BERT\textsc{-65k}       & DNN & 99.92 & 99.97 & 99.85 & 96.10 & 98.09 & 94.36 & 0 \\
    text-embedding-3        & DNN & 99.82 & 99.97 & 99.63 & 89.75 & 97.11 & 83.29 & 0\\
  \bottomrule
\end{tabular}
\end{center}
\end{table*}

\end{document}